\documentclass[10pt, twocolumn]{IEEEtran}

\IEEEoverridecommandlockouts

\usepackage{graphicx}
\usepackage{subfigure}
\usepackage{color}
\usepackage{epsfig}
\usepackage{amssymb}
\usepackage{amsmath}
\usepackage{amsthm}
\usepackage{latexsym}
\usepackage{mathrsfs}

\newcommand{\ds}{\displaystyle}
\newcommand{\bPP}[1]{{\mathrm{P}}_{#1}}
\newcommand{\bPr}[1]{\mathrm{P}\left(#1\right)}

\newcommand{\tPP}[1]{{\tilde{\mathrm{P}}}_{#1}}
\newcommand{\tP}[2]{\tilde{\mathrm{P}}_{#1}\left({#2}\right)}
\newcommand{\tPs}[3]{\tilde{\mathrm{P}}^{#1}_{#2}\left({#3}\right)}
\newcommand{\bP}[2]{{\mathrm{P}}_{#1}\left({#2}\right)}
\newcommand{\bE}[2]{{\mathbb{E}}_{#1}\left[{#2}\right]}
\newcommand{\bEE}[1]{{\mathbb{E}}\left[{#1}\right]}
\newcommand{\cX}{{\mathcal X}}
\newcommand{\cY}{{\mathcal Y}}
\newcommand{\cA}{{\mathcal A}}
\newcommand{\cZ}{{\mathcal Z}}
\newcommand{\cB}{{\mathcal B}}
\newcommand{\cM}{{\mathcal M}}
\newcommand{\cN}{{\mathcal N}}
\newcommand{\cF}{{\mathscr F}}

\newcommand{\cU}{{\mathcal U}}

\newcommand{\cV}{{\mathcal V}}
\newcommand{\cT}{{\mathcal T}}

\newcommand{\cL}{{\mathcal L}}
\newcommand{\cI}{{\mathcal I}}
\newcommand{\bl}{{\overline{\lambda}}}
\newcommand{\ls}{{\lambda_{\mathtt{sum}}}}

\newcommand{\bF}{{\mathbf{F}}}
\newcommand{\bi}{{\mathbf{i}}}

\newcommand{\ep}{\epsilon}
\newcommand{\la}{\lambda}

\newtheorem{theorem}{Theorem}
\newtheorem{proposition}[theorem]{Proposition}
\newtheorem*{corollary}{Corollary}

\newtheorem{lemma}[theorem]{Lemma}

\theoremstyle{remark}
\newtheorem*{remark*}{Remark}
\newtheorem*{remarks*}{Remarks}

\theoremstyle{definition}
\newtheorem{definition}{Definition}

\begin{document}

\title{How Many Queries Will Resolve Common Randomness?}
\author{Himanshu Tyagi and Prakash Narayan$^\dag$}
\maketitle {\renewcommand{\thefootnote}{}\footnotetext{
\vspace{.02in}

\noindent 
$^\dag$H. Tyagi and P. Narayan are with the Department of
Electrical and Computer Engineering and the Institute for Systems
Research, University of Maryland, College Park, MD 20742, USA.
E-mail: \{tyagi, prakash\}@umd.edu.

\noindent This work was supported by the U.S.
National Science Foundation under Grants CCF0830697 and
CCF1117546.

A version of this paper will be presented at the IEEE International
 Symposium on Information Theory, Istanbul, Turkey, July 7 - July 12, 2013.
 }

}

\begin{abstract}
A set of $m$ terminals, observing 
correlated signals, communicate interactively to 
generate common randomness for a given subset of them. 
Knowing only the communication, how many
direct queries of the value of the common randomness
will resolve it? A general upper bound, valid for arbitrary 
signal alphabets, is developed for the number of such queries
by using a query strategy that applies to all common randomness
and associated communication. When the underlying signals
are independent and identically distributed repetitions of $m$
correlated random variables, the number of queries can be exponential
in signal length. For this case, the mentioned upper bound is
tight and leads to a single-letter formula for the largest query 
exponent, 
which coincides with the secret key capacity of a corresponding 
multiterminal source model. In fact, the upper bound constitutes
a strong converse for the optimum query exponent, and implies also
a new strong converse for secret key capacity. A key tool, 
estimating the size of a large probability set in terms of R\'enyi
entropy, is interpreted separately, too, as a lossless block coding 
result for general sources. As a particularization, it yields the classic result 
for a discrete memoryless source.
\end{abstract}

%
\section{Introduction}

A set of terminals observing correlated signals agree on
common randomness (CR), i.e., shared bits, by communicating
interactively among themselves. What is the maximum number
of queries of the form ``Is  CR $= l$?" with yes-no answers, that 
an observer of (only) the communication must ask in order to resolve the 
value of the CR? As an illustration, suppose that two terminals observe, 
respectively, $n$ independent and identically distributed (i.i.d.) 
repetitions of the finite-valued random variables (rvs) $X_1$
and $X_2$. The terminals agree on CR $X_1^n$ with terminal $1$
communicating to terminal $2$ a Slepian-Wolf codeword of rate 
$H\left(X_1\mid X_2\right)$ obtained by random binning. 
An observer of the bin index can ascertain
the value of CR  with large probability in approximately 
$\exp\left[nI\left(X_1\wedge X_2\right)\right]$ queries 
(corresponding to bin size). Our results show
that more queries cannot be incurred by any other form of CR and 
associated interactive communication.

In a general setting, terminals $1, ..., m$ observe, respectively,
$n$ i.i.d. repetitions of the rvs $X_1, ..., X_m$, and communicate
interactively to create CR, say $L$, for the terminals in a given
subset $\cA \subseteq \{1, ..., m\}$. For appropriate 
CR $L$ and communication $\bF$, the number of queries of the form 
``Is  $L = l$?" that an observer of $\bF$ must ask to resolve $L$ is 
exponential in $n$. We find a single-letter formula for the largest
exponent $E^*$. Remarkably, this formula coincides with the secret 
key (SK) capacity for a multitermial source model with underlying
rvs $X_1, ..., X_m$ \cite{CsiNar04, CsiNar08}. The latter is the largest
rate of nearly uniformly distributed CR for $\cA$ that meets the 
security requirement of being nearly independent
of the communication used to generate it. 
While it is to be expected that $E^*$ is no smaller than SK
capacity, the less-restricted $E^*$ may seem {\it a priori}
to be larger. But it is not so.
The coincidence brings out, in effect,
an equivalence between inflicting a maximum number of queries on an 
observer of $\bF$ on the one hand, and imposing the explicit secrecy
constraint above on the other hand. In fact, as in the achievability proof of
SK capacity in \cite{CsiNar04}, the exponent $E^*$ is achieved by the 
terminals in $\cA$ attaining ``omniscience," i.e., by generating CR
$L = \left(X_1^n, ..., X_m^n\right)$ for $\cA$, 
using communication $\bF$ of minimum rate.

Alternatively, $E^*$ can be interpreted as the smallest rate of a list of CR
values produced by an observer of $\bF$ which contains $L$ with large probability.

Our main contribution is a new technique for proving 
converse results involving CR with interactive communication.
It relies on query strategies for $L$ given $\bF$ that do not 
depend explicitly on the form of $L$ or $\bF$, and do not 
require the rvs $\left(X_{1t}, ..., X_{mt}\right)_{t=1}^n$ to be
finite-valued or i.i.d. 
In fact, our converse results hold even when 
the underlying alphabets are arbitrary, but under 
mild technical assumptions. Jointly Gaussian rvs are treated
as a special case.
Furthermore, our converses are strong 
in that the characterization of $E^*$ does not depend on the
probability of recovery of the CR. This, in turn, leads to 
a new strong converse result for the
SK capacity of the multiterminal source model \cite{CsiNar04}, \cite{CsiNar08}. A byproduct 
of our technique is a simple lossless block coding result 
for general finite sources, in terms of R\'enyi entropies. A
particularization recovers the classic lossless block coding 
result for i.i.d. sources \cite{Sha48} without recourse to the
asymptotic equipartition property (AEP).

The number of queries above can be interpreted as a measure of
the correlation among the random signals observed by the terminals: 
A stronger correlation
necessitates more queries for resolving the CR that can be generated
by them. Such a measure 
of correlation is in the spirit of the body of work on 
``guessing" 
the value of an rv based on a correlated observation
\cite{Mas94, Ari96, AriMer99, HanSun11}. 

The problem formulation and our main result characterizing 
the optimum query exponent  are given in the 
next section. Simple and essential technical tools which also 
may be of independent interest are presented in Section \ref{s:tools}. 
Achievability is proved in Section \ref{s:achievability}. The less 
complex converse proof for the case $\cA =\{1, ..., m\}$ 
is given in Section \ref{s:converse_A=M}. However, this proof
does not extend to an arbitrary $\cA\subseteq \{1, ..., m\}$, for which a 
different converse is provided in Section \ref{s:converse_A<M}. 
Section \ref{s:strong_converse} contains the strong converse result
for SK capacity. A converse for the optimum query exponent 
for rvs with arbitrary alphabets is
proved in Section \ref{s:general_converse}, with jointly
Gaussian rvs as a special case. The discussion in Section \ref{s:Discussion}
includes the mentioned lossless block coding result for general sources.

\section{Main Result}\label{s:Main Results}

Let $X_1, \ldots, X_m$, $m \geq 2$, be rvs with finite alphabets
$\cX_1, \ldots, \cX_m$, respectively, and with a known joint
probability mass function (pmf) $\bPP{X_1, ..., X_m}$. 
For any nonempty set $\cA \subseteq
\cM = \{1, \ldots, m\}$, we denote $X_\cA = (X_i,\ i \in \cA)$.
We denote $n$ i.i.d. repetitions of $X_{\cM} =
(X_1, \ldots, X_m)$ with values in $\cX_\cM = \cX_1 \times \ldots
\times \cX_m$ by $X_{\cM}^n = (X_1^n, \ldots, X_m^n)$ with values
in $\cX_\cM^n = \cX_1^n\times \ldots \times \cX_m^n$. Given 
$\epsilon > 0$, for rvs $U, V,$ we say that
$U$ is $\epsilon$-\emph{recoverable} from $V$ if $\bPr{ U \neq
f(V)} \leq \epsilon$ for some function $f(V)$ of $V$. The cardinality 
of the range of the rv $U$ is denoted by $\|U\|$, and the complement
of a set  $A$ by $A^c$. All
logarithms and exponentials are with respect to the base $2$.

We consider a multiterminal source model for 
generating CR using interactive communication.
Terminals $1, \dots, m$ observe, respectively, the
sequences $X_1^n, \ldots, X_m^n$, of length $n$.  
The terminals in a given set
$\cA \subseteq \cM$ wish to
generate CR using communication over a noiseless
channel, possibly interactively in several rounds. 
\begin{definition}\label{d:PubComm}
Assume without any loss of generality that the communication of
the terminals in $\cM$ occurs in consecutive time slots in $r$
rounds, where $r$ can depend on $n$ but is finite for every $n$.
Such communication is described in terms of the mappings
\begin{align}\nonumber
f_{11},\ldots,f_{1m},
f_{21},\ldots,f_{2m},\ldots,f_{r1},\ldots,f_{rm},
\end{align}
with $f_{ji}$ corresponding to a message in time slot $j$ by
terminal $i$, $1 \leq j \leq r$, $1 \leq i \leq m$; in general,
$f_{ji}$ is allowed to yield any function of $X_i^n$ and of
previous communication 
$$\phi_{ji} = \{ f_{kl}: k<j,\ l
\in \cM \ \text{or} \ k = j,\  l< i\}$$
The corresponding rvs
are termed collectively as {\it interactive communication}
\begin{align}\nonumber
\bF = \{F_{11},\ldots,F_{1m},
F_{21},\ldots,F_{2m},\ldots,F_{r1},\ldots,F_{rm}\},
\end{align}
where $\bF = \bF^{(n)}(X^n_\cM)$; the rv corresponding 
to $\phi_{ji}$ is denoted by $\Phi_{ji}$. Local randomization at the terminals
is not considered here for ease of exposition. In fact, allowing such
randomization does not improve our result; see Section \ref{s:Discussion}-B.
\end{definition}
\begin{definition}\label{d:CR_def}
Given interactive communication $\bF$ as above,
an rv $L = L^{(n)}\left(X_\cM^n\right)$ is \emph{$\ep$-common 
randomness} ($\ep$-CR) for $\cA$ from $\mathbf{F}$
if it is $\ep$-recoverable from $\left(X_i^n, \bF\right)$, $i \in \cA$, i.e., 
if there exist rvs $L_i = L_i^{(n)}\left(X_i^n, \mathbf{F}\right)$, $i \in \cA$,
satisfying 
\begin{align}
\bPr{L_i = L,\,\, i \in \cA} \geq 1 - \ep.
\label{e:CR_recover}
\end{align}
The rv $L_i$ will be called an estimate of $L$ at terminal $i \in \cA$.
\end{definition}
A querier observing the communication $\bF$ wants to resolve 
the value of this CR $L$ by asking questions of the form ``Is $L=l$?"
with yes-no
answers. While queries of this form have been termed ``guessing" 
\cite{Mas94, Ari96, AriMer99, HanSun11}, we use the terminology 
``query" since our approach covers a broader class of query strategies;
see Section \ref{s:Discussion}-B.
\begin{definition}
For rvs $U, V$ with values in the sets $\cU, \cV$, a 
\emph{query strategy} $q$ for $U$ given $V=v$ is a 
bijection $q(\cdot| v): \cU \rightarrow \{1, ..., |\cU|\}$, where 
the querier, upon observing $V=v$, asks the question
``Is $U=u$?" in the $q(u|v)^{\text{th}}$ query.
\end{definition}
Thus, a query strategy $q$ for resolving a CR $L$ on 
the basis of an observed communication $\bF = \bi$
 is an ordering of the possible values of $L$. The terminals 
 seek to generate a CR $L$ for $\cA$ using communication
 $\bF$ so as to make the task of the querier observing $\bF$
 as onerous as possible. For instance, if $L$ were to be 
 independent of $\bF$, then the querier necessarily must search 
 exhaustively over the set of possible values of $L$, which 
can be exponentially large (in $n$).
 \begin{definition}\label{d:query_exponent}
 Given  $0<\ep <1$, a \emph{query exponent} $E>0$ is 
 $\ep$-achievable if for every $0< \ep^\prime <1$, 
there exists an $\ep$-CR 
 $L =  L^{(n)}\left(X_\cM^n\right)$ for $\cA \subseteq \cM$ 
 from communication 
 $\bF = \bF\left(X_\cM^n\right)$ such that for every query
 strategy $q$ for $L$ given $\bF$,
 \begin{align}
\mathrm{P}\big(q(L\mid \bF) \geq \exp (nE)\big) > 1-\ep^\prime,\label{e:query_exp}
 \end{align}
 for all $n \geq N(\ep, \ep^\prime)$.
The $\ep$-optimum query exponent, denoted $E^*(\ep)$, is
 the supremum of all $\ep$-achievable 
query exponents; $E^*(\ep)$ is nondecreasing in $\ep$.
The \emph{optimum query exponent} $E^*$  is the infimum 
of $E^*(\ep)$ for $0< \ep < 1$, i.e.,
$$E^* = \lim_{\ep \rightarrow 0}E^*(\ep).$$
 \end{definition}
 \begin{remark*}
 Clearly, $0 \leq  E^* \leq \log|\cX_\cM|$.
 \end{remark*}
 
\noindent Condition (\ref{e:query_exp}) forces any query strategy adopted 
by the querier to have an exponential complexity (in $n$) with large 
probability;
$E^*$ is the largest value of the exponent that can be inflicted on the
querier.

Our main result is a single-letter characterization of the optimum query
exponent $E^*$. 
Let 
\begin{align}
\cB = \left\{B\subsetneq \cM :  B \neq \emptyset, \cA\nsubseteq B\right\}.
\label{e:set_B}
\end{align} 
 Let $\Lambda(\cA)$ be the set
 of all collections 
 $\lambda = \left\{\lambda_B : B\in \cB\right\}$ of weights $0\leq \lambda_B \leq 1$, 
 satisfying
 \begin{align}
\displaystyle\sum_{B  \in \cB: B\ni i} 
\lambda_B = 1, \quad i \in \cM.\label{e:Lambda(A)}
 \end{align}
 Every $\lambda \in \Lambda(\cA)$ is called a {\it fractional partition} 
 of $\cM$ (see \cite{CsiNar08, MadBar07, MadMarTet12, MadTet10}).
 \begin{theorem}\label{t:main}
 The optimum query exponent $E^*$ equals 
 \begin{align}
E^* = E^*(\ep) = H\left(X_\cM\right) - &\max_{\la \in \Lambda(\cA)} 
\sum_{B\in \cB }\lambda_B 
H\left(X_B\mid X_{B^c}\right), 
\nonumber
\\&\hspace*{2.6cm} 0<\ep< 1.\label{e:E*_char}
 \end{align}
 \end{theorem}
 Remarkably, the value of $E^*$ coincides with the \emph{secret key} (SK)
 capacity of a multiterminal source model \cite{CsiNar04, CsiNar08}. The
 latter is the largest rate of a CR $K = K\left(X_\cM^n\right)$
 for $\cA$ from communication $\bF$, with 
 $K$
satisfying the ``secrecy constraint" of \cite{CsiNar04}:
 \begin{align}
\lim_{n} s_{in}(K;\bF)= 0,\label{e:SK_security}
 \end{align}
where
the security index $s_{in}$
is given by
\begin{align}
s_{in}(K;\bF) =  \log\|K\| - H(K\mid \bF) = D\left(\bPP{K, \bF}\, \|\, \bPP{\mathrm{unif}}\times \bPP{\bF}\right),
\label{e:s_in_def}
\end{align}
with $\bPP{\mathrm{unif}}$ being the uniform pmf on 
$\{1, ..., \|K\|\}$.
In fact, the achievability proof of Theorem \ref{t:main} is
 straightforward and employs, in effect, an SK in forming 
an appropriate CR $L$. We show that for such
 a CR $L$, any query strategy is tantamount to an 
 exhaustive search over the set of values of the SK, a feature
 that is apparent for a ``perfect" SK with $I(K \wedge \bF)= 0$.
 The difficult step in the proof of Theorem \ref{t:main} is the 
 converse part which involves an appropriate query
 strategy, for arbitrary $L$ and $\bF$, that limits the incurred 
 query 
 exponents. 
Our \emph{strong} converse
yields a uniform upper bound for $E^*(\ep)$, $0< \ep <1$.

We shall see that while the expression for $E^*$ in (\ref{e:E*_char}) lends 
itself to the achievability proof of 
Theorem \ref{t:main} in Section \ref{s:achievability}, alternative forms are suited better 
for the converse proof. For the latter, denoting 
\begin{align}
\ls =\sum_{B \in \cB} \la_B,
\label{e:lambda_sum}
\end{align}
the expression (\ref{e:E*_char}) can be written also as
\begin{align}
E^* = \min_{\la \in \Lambda(\cA)} \left[\sum_{B \in \cB} \la_B H\left(X_{B^c}\right) - \left(\ls - 1\right)H\left(X_\cM\right)\right],
\label{e:E*_alt1}
\end{align}
which is used in the converse proof for 
an arbitrary $\cA\subseteq \cM$ in Section \ref{s:converse_A<M}. 
The converse proof for the 
case $\cA =\cM$ is facilitated by the fact that the 
right-side of (\ref{e:E*_alt1}) can be expressed equivalently as \cite{ChaZhe10} (see also
\cite[Example 4]{CsiNar04})
\begin{align}
\min_{\pi} \frac{1}{|\pi| - 1} D\bigg(\bPP{X_\cM} \| \prod_{i=1}^{|\pi|}\bPP{X_{\pi_i}} \bigg),
\label{e:C_alternate_expression}
\end{align}
where the minimum is over all (nontrivial) partitions $\pi=\left(\pi_1, ..., \pi_k\right)$
of $\cM$ with $|\pi| = k$ parts, $2\leq k \leq m$. 

\section{Technical Tools}\label{s:tools}

The following simple observation relates the number of queries in a query strategy $q$
to the cardinality of an associated set.

\begin{proposition}\label{p:query_cardinality}
Let $q$ be a query strategy for $U$ given $V=v$, $v \in \cV$. Then,
\begin{align}\nonumber
|\{u\in \cU: q(u| v) \leq \gamma\}| \leq \gamma.
\end{align}
\end{proposition}
{\it Proof.} The claim is straightforward since $q(\cdot| v)$ is a bijection.\qed

\noindent For rvs $U, V$, finding a lower bound for $q(U| V)$ involves finding a suitable
upper bound for the conditional probabilities $\bP{U| V}{\cdot \mid \cdot}$. This idea is formalized by the following lemma.

\begin{lemma}\label{l:equivalence}
Given $\gamma > 0$ and $0< \delta <1/2$, let the rvs $U, V$, 
satisfy 
\begin{align}
\bPr{\left\{(u,v): \bP{U| V}{u| v} \leq \frac{\delta}{\gamma}\right\}}
\geq 1 - \delta.\label{e:ub1_condP}
\end{align}
Then for every query strategy $q$ for $U$ given $V$,
\begin{align}
\bPr{ q(U| V) \geq \gamma}\geq 1 - \ep^\prime,
\label{e:lb_query}
\end{align}
for all $ \ep^\prime \geq 2\delta$.

Conversely, if (\ref{e:lb_query}) holds for every query 
strategy $q$ for $U$ given $V$, with $0 <  \ep^\prime \leq (1 - \sqrt{\delta})^2$,
then
\begin{align}
\bPr{\left\{(u,v): \bP{U| V}{u| v} \leq \frac{1}{\gamma}\right\}} \geq \delta.
\label{e:ub2_condP}
\end{align}
\end{lemma}
{\it Proof.} Suppose (\ref{e:ub1_condP}) holds but not (\ref{e:lb_query}). Then
there exists $q$ with 
\begin{align}
\bPr{q(U| V) < \gamma} > \ep^\prime.\label{e:l1_1}
\end{align}
From (\ref{e:ub1_condP}) and (\ref{e:l1_1})
\begin{align}
&\bPr{\left\{(u, v): \bP{U| V}{u| v} \leq \frac{\delta}{\gamma}, \,\,
q(u| v) < \gamma\right\}}
\nonumber
\\&> 1-\delta + \ep^\prime - 1 = \ep^\prime - \delta.
\label{e:l1_2}
\end{align}
On the other hand, the left side of (\ref{e:l1_2}) equals
\begin{align}
&\sum_{v}\bP{V}{v}\sum_{u: q(u| v) < \gamma,\,\, \bP{U| V}{u| v} \leq 
\frac{\delta}{\gamma}}\bP{U| V}{u| v}
\nonumber
\\&\leq \gamma. \frac{\delta}{\gamma},
\qquad \text{by Proposition \ref{p:query_cardinality}}
\nonumber
\\&=\delta,
\nonumber
\end{align}
which contradicts (\ref{e:l1_2}) since $\ep^\prime \geq 2\delta$.

For the converse, suppose that 
(\ref{e:ub2_condP}) does not hold; then, we 
show that a query strategy $q_0$ exists which violates
(\ref{e:lb_query}) when $0 < \ep^\prime \leq (1 - \sqrt{\delta})^2$.
The negation of (\ref{e:ub2_condP}) is
\begin{align}
\bPr{\left\{(u,v): \bP{U| V}{u| v} > \frac{1}{\gamma}\right\}} > 1 - \delta,
\nonumber
\end{align}
which, by a reverse Markov inequality\footnote{The reverse
Markov inequality states that for rvs $U, V$ with $\bPr{(U, V) \in S} \geq 1- \ep$ for some
$S\subseteq \cU \times \cV$, there exists $\cV_0\subseteq \cV$ such that
 $\bPr{(U, V) \in S \mid V = v} \geq 1- \sqrt{\ep}$, $v \in \cV_0$, and $\bPr{V\in \cV_0} \geq 1- \sqrt{\ep}$.} \cite[p. 157]{Loe60} (see also \cite[p. 153]{GacKor73}),
gives a set $\cV_0 \subseteq \cV$ with
\begin{align}
\bP{V}{\cV_0} > 1 - \sqrt{\delta},
\label{e:l1_3}
\end{align}
and 
\begin{align}
\bP{U| V}{\left\{(u: \bP{U| V}{u| v} > \frac{1}{\gamma}\right\} \,\bigg\vert\, v} > 1 - \sqrt{\delta}, \quad v\in \cV_0.
\label{e:l1_4}
\end{align}
Denoting by $\cU_v$ the set $\{\cdot\}$ in (\ref{e:l1_4}), we have 
\begin{align}
1\geq \bP{U| V}{\cU_v\mid v} > \frac{|\cU_v|}{\gamma},
\nonumber
\end{align}
so that
\begin{align}
|\cU_v| < \gamma, \quad v\in \cV_0.
\label{e:l1_5}
\end{align}
For each $v\in \cV_0$, order the elements of $\cU$ arbitrarily but with
the first $|\cU_v|$ elements being from $\cU_v$. This ordering 
defines a query strategy $q_0(\cdot | v)$, $v \in \cV_0$; for $v\notin \cV_0$,
let $q_0(\cdot| v)$ be defined arbitrarily. Then for $v \in \cV_0$, $u \in \cU_v$,
\begin{align}
q_0(u| v) < \gamma
\nonumber
\end{align}
by (\ref{e:l1_5}), so that
\begin{align}
\bPr{q_0(U| V) < \gamma} &\geq 
\sum_{v\in \cV_0}\sum_{u \in \cU_v}\bP{U,V}{u, v}
\nonumber
\\&> (1 - \sqrt{\delta})^2,
\label{e:q_0_ub}
\end{align}
by (\ref{e:l1_3}) and (\ref{e:l1_4}).
So, $q=q_0$ violates (\ref{e:lb_query}) when $\ep^\prime \leq (1-\sqrt{\delta})^2$.\qed

The next result relates the cardinalities of large probability sets to R{\'e}nyi entropy.
The first part is used in the converse proofs of Theorem \ref{t:main}. The mentioned
result is of independent interest. For instance, in Section \ref{s:Discussion} it is shown 
to yield an elementary alternative proof of the source coding theorem for an i.i.d.
(finite-valued) source.

\begin{definition} \cite{Ren61}\label{d: Renyi_Entropy}
Let $\mu$ be a nonnegative measure on $\cU$. For $0 \leq \alpha \neq 1$, the 
\emph{R\'enyi entropy of order $\alpha$} of $\mu$ is defined as 
\begin{align}
H_\alpha(\mu) = \frac{1}{1-\alpha}\log \sum_{u \in \cU}\mu(u)^\alpha.
\nonumber
\end{align}
\end{definition}
\begin{lemma}\label{l:card_largePset}
(i) For every $0 < \delta < \mu(\cU)$, there exists a set
$\cU_\delta \subseteq \cU$ such that 
\begin{align}
\mu\left(\cU_\delta\right) \geq \mu(\cU) -\delta,
\label{e:l2_1}
\end{align}
and 
\begin{align}
|\cU_\delta| \leq \delta^{-\alpha/ (1 - \alpha)} \exp\left(H_\alpha(\mu)\right), 
\qquad 0\leq \alpha <1.
\label{e:ub_card_largePset}
\end{align}

\noindent (ii) Conversely, for $\delta, \delta^\prime >0$, $\delta + \delta^\prime <\mu(\cU)$,
any set $\cU_\delta \subseteq \cU$ with $\mu\left(\cU_\delta\right)$ as in (\ref{e:l2_1})
must satisfy
\begin{align}
|\cU_\delta| \geq \left(\delta^\prime\right)^{1/ (\alpha- 1)} (\mu(\cU) - \delta -\delta^\prime)\exp\left(H_\alpha(\mu)\right), \qquad \alpha> 1.
\label{e:lb_card_largePset}
\end{align}
\end{lemma}
{\it Proof.} 
(i) For $0 \leq \alpha <1 $, defining 
$$\cU_\delta = \left\{u\in \cU: \mu(u) > \delta^{\frac{1}{1-\alpha}}\exp\left[-H_\alpha(\mu)\right]\right\},$$
we get 
\begin{align}
\mu(\cU) &= \mu(\cU_\delta) + \sum_{u:\,\, \mu(u) \,\leq\, \delta^{\frac{1}{1-\alpha}}\exp\left[-H_\alpha(\mu)\right]}\mu(u).
\nonumber
\end{align}
Writing the summand in the right-side above as 
$\mu(u) = \mu(u)^\alpha\mu(u)^{1-\alpha}$, we obtain
\begin{align}
\mu(\cU) &\leq \mu(\cU_\delta) +  \delta\exp\left[-(1-\alpha)H_\alpha(\mu)\right]
\sum_{u \in \cU} \mu(u)^\alpha
\nonumber\\
&= \mu\left(\cU_\delta\right) + \delta,
\nonumber
\end{align}
which is (\ref{e:l2_1}). Furthermore, 
\begin{align}
\exp\left[(1-\alpha)H_\alpha(\mu)\right]
&= \sum_{u \in \cU} \mu(u)^\alpha
\nonumber\\ 
&\geq \sum_{u\in \cU_\delta}\mu(u)^\alpha 
\nonumber\\
&\geq  |\cU_\delta|\delta^\frac{\alpha}{1-\alpha}\exp\left[-\alpha H_\alpha(\mu)\right],
\nonumber
\end{align}
which gives (\ref{e:ub_card_largePset}).

\noindent (ii) By following the steps in the proof of (i), for $\alpha >1$, it can shown that
the set 
\begin{align}
\cU_0 = \left\{u \in \cU: \mu(u) < \left(\delta^\prime\right)^{1/  (1-\alpha)}
\exp[- H_\alpha(\mu)]\right\}
\label{e:def_U_0}
\end{align}
has 
$$\mu(\cU_0) > \mu(\cU) - \delta^\prime,$$
which, with (\ref{e:l2_1}), gives
$$\mu(\cU_0\cap \cU_\delta) > \mu(\cU) - \delta - \delta^\prime.$$ 
Since by (\ref{e:def_U_0})
\begin{align}
\mu(\cU_0\cap \cU_\delta) < \left|\cU_0 \cap \cU_\delta\right|\left(\delta^\prime\right)^{1/  (1-\alpha)}\exp[- H_\alpha(\mu)],
\nonumber
\end{align}
(\ref{e:lb_card_largePset}) follows.
\qed

Finally, the following simple observation will be useful.
\begin{proposition}\label{p:measure_change}
For pmfs $Q_1, Q_2$, on $\cV$,
\begin{align}
Q_1\left(\left\{v: Q_1(v) \geq \delta Q_2(v)\right\}\right) \geq 1 - \delta, \qquad 
0< \delta <1.
\nonumber
\end{align}
\end{proposition}
{\it Proof.} The claim follows from 
\begin{align}
\sum_{v\in \cV: Q_1(v) < \delta Q_2(v)}Q_1(v) < \sum_{v\in \cV: Q_1(v) < \delta Q_2(v)}\delta\,  Q_2(v) \leq \delta.
\nonumber
\end{align}
\qed
 
 \section{Achievability proof of Theorem \ref{t:main}}\label{s:achievability}

Denoting the right-side of (\ref{e:E*_char}) by C, we claim,
for $0< \ep <1$, $0 < \delta < 1/   2$, $\beta > 0$, the existence 
of an $\ep$-CR $L = X_\cM^n$ for $\cA$ from $\mathbf{F}$
with 
\begin{align}
\bPr{\left\{ \left(x_\cM^n, \bi\right): \bP{L \mid \mathbf{F}}{x_\cM^n\mid \bi}
\leq \delta \exp\left[-n(C - \beta)\right]
\right\}} \geq 1 - \delta,
\label{e:t1_1}
\end{align}
 for all $n$ sufficiently large. Then the assertion of the theorem follows by
 applying the first part of Lemma \ref{l:equivalence} with
 $U = L$, $V= \mathbf{F}, \gamma= \exp[n(C-\beta)]$, to conclude from (\ref{e:lb_query})
 that 
 $$E^*(\ep) \geq C,$$
 since $\beta> 0$ was chosen arbitrarily.
 
 Turning to the mentioned claim, it is shown in \cite[Proposition 1]{CsiNar04}, \cite[Theorem 3.1]{CsiNar08} that there exists communication $\mathbf{F}$
 such that $L = X_\cM^n$ is $\ep$-CR for $\cA$ from $\mathbf{F}$ with 
 \begin{align}
 \frac{1}{n}\log\|\bF\| \leq \max_{\lambda \in \Lambda(\cA)} 
\sum_{B\in \cB }\lambda_B 
H\left(X_B\mid X_{B^c}\right) + \frac{\beta}{3},
 \label{e:rate_RCO}
 \end{align}
 for all $n$ sufficiently large. Using Proposition \ref{p:measure_change}
 with $Q_1 = P_\mathbf{F}$ and $Q_2$ being the uniform pmf over the 
 range of $\mathbf{F}$, we get 
 \begin{align}
 \bP{\mathbf{F}}{\left\{\bi: \bP{\mathbf{F}}{\bi} \geq \frac{\delta}{2\|\mathbf{F}\|}\right\}}
 \geq 1 - \frac{\delta}{2}.
 \label{e:t1_2}
 \end{align}
 Also, for $x_\cM^n$ in the set $\mathcal{T}_n$ of $\bPP{X_\cM}$-typical sequences
 with constant $\delta$ \cite[Definition 2.8]{CsiKor11}, we have 
 \begin{align}
 \bP{X_\cM^n}{x_\cM^n} \leq \exp\left[-n\left(H\left(X_\cM\right) - \frac{\beta}{3}\right)\right]
 \label{e:t1_3}
 \end{align}
 and 
  \begin{align}
 \bP{X_\cM^n}{\mathcal{T}_n} \geq 1 - \frac{\delta}{2},
\nonumber
 \end{align}
 for all $n$ sufficiently large. Denoting by $\mathcal{I}_0$ the set on the left-side of (\ref{e:t1_2}), it follows that 
\begin{align}
\bPr{X_\cM^n \in \mathcal{T}_n, \mathbf{F}\in \mathcal{I}_0} \geq 1 - \delta.
\label{e:t1_4}
\end{align}
The claim results from (\ref{e:t1_2})-(\ref{e:t1_4}) upon observing that for $\left(x_\cM^n, \bi\right) \in \mathcal{T}^n\times \mathcal{I}_0$,
\begin{align}
\bP{X_\cM^n\mid \mathbf{F}}{x_\cM^n\mid \bi} &= \frac{\bP{X_\cM^n}{\left(x_\cM^n\right)}\mathbf{1}
\left(\mathbf{F}\left(x_\cM^n\right)=\bi\right)}{\bP{\mathbf{F}}{\bi}}
\nonumber
\\&\leq \frac{2 \exp\left[-n\left(H\left(X_\cM\right) - \frac{\beta}{3}\right)\right]
\|\mathbf{F}\|}{\delta}\nonumber
\\&\leq \delta\exp[-n(C-\beta)],
\nonumber
\end{align}
for all $n$ large enough, where the last inequality is by (\ref{e:rate_RCO}).\qed

\begin{remark*}
The achievability proof brings out a connection between a large probability
uniform upper bound $\kappa$ for $\bPP{L}$, the size $\|\mathbf{F}\|$
of the communication $\mathbf{F}$, and the associated number of queries needed.
Loosely speaking, the number of queries is approximately $\frac{1}{\|\mathbf{F}\|\kappa}$, which reduces to $\frac{\|L\|}{\|\mathbf{F}\|}$ if $L$ is nearly uniformly
distributed.
\end{remark*}

\section{Converse proof of Theorem \ref{t:main} for $\cA = \cM$}\label{s:converse_A=M}
Recalling the expression for $E^*$ in (\ref{e:C_alternate_expression}), given a partition $\pi$ of $\cM$ with $|\pi|=k$, $2\leq k \leq m$, we
observe that for a consolidated source model with $k$ sources and underlying rvs $Y_1, ..., Y_k$ where\footnote{ For specificity, the elements in each $\pi_i$ are arranged in 
increasing order.} $Y_i = X_{\pi_i}$, the $\ep$-optimum query exponent $E^*_\pi(\ep)$ can be no smaller than $E^*(\ep)$ (since the terminals in each $\pi_i$ coalesce, in effect).
\begin{theorem}\label{t:converse}
For every partition $\pi$ of $\cM$ with $|\pi| = k$, 
\begin{align}
E^*_\pi(\ep) \leq \frac{1}{k - 1} D\bigg(\bPP{Y_1, ..., Y_k} \| \prod_{i=1}^k\bPP{Y_i} \bigg), \qquad 0< \ep < 1,
\nonumber
\end{align}
and so
\begin{align}
E^*(\ep) \leq \min_{\pi} E^*_\pi(\ep) \leq \min_{\pi} \frac{1}{|\pi| - 1} D\bigg(\bPP{X_\cM} \| \prod_{i=1}^{|\pi|}\bPP{X_{\pi_i}} \bigg).
\nonumber
\end{align}
\end{theorem}
\noindent Theorem \ref{t:converse} establishes, in view of (\ref{e:C_alternate_expression}), 
the converse part of Theorem \ref{t:main}  when 
$\cA =\cM$.

The proof of Theorem \ref{t:converse} relies on the following
general result, which holds for queries of CR generated in a 
multiterminal source model with underlying rvs 
$Y_1, ..., Y_k$ for $n=1$. 
\begin{theorem}\label{t:general_converse}
Let $L = L\left(Y_1, ..., Y_k\right)$ be $\ep$-CR for $\{1, ..., k\}$ 
from interactive communication 
$\mathbf{F} = \mathbf{F}\left(Y_1, ..., Y_k\right)$, $0< \ep < 1$.
Given $\delta> 0$ such that $\delta + \sqrt{\delta + \ep} <1$, let $\theta$ be such that 
\begin{align}
\bPr{\left\{ \left(y_1, ...,y_k\right): \frac{\bP{Y_1, ..., Y_k}{y_1, ..., y_k}}{\prod_{i=1}^k\bP{Y_i}{y_i}}  \leq \theta \right\}}\geq 1 - \delta.
\label{e:Entropy_boundsOnP}
\end{align}
Then, there exists a query strategy $q_0$ for $L$ given $\mathbf{F}$
such that
\begin{align}
\bPr{q_0(L\mid \mathbf{F} ) \leq  \left(\frac{\theta}{\delta^2}\right)^{\frac{1}{k-1}}}  \geq (1 - \delta	- \sqrt{\delta + \ep})^2.
\label{e:q_adversary}
\end{align}
\end{theorem}
{\it Proof of Theorem \ref{t:converse}.} We apply Theorem \ref{t:general_converse} to
$n$ i.i.d. repetitions of the rvs $Y_1, ..., Y_k$. Denoting by $\mathcal{T}_n^\prime$ the set of $\bPP{Y_1, ..., Y_k}$-typical sequences with constant $\delta$, we have 
\begin{align}
\bP{Y_1^n, ..., Y_k^n}{\mathcal{T}^\prime_n}\geq 1 - \delta, 
\nonumber
\end{align}
and for $\left(y_1^n, ..., y_k^n\right) \in \mathcal{T}^\prime_n$,
\begin{align}
&\frac{\bP{Y_1^n, ..., Y_k^n}{y_1^n, ..., y_k^n}}{\prod_{i=1}^k\bP{Y_i^n}{y_i^n}} 
\nonumber
\\&\leq\exp\bigg[n\bigg( \sum_{i=1}^k H\left(Y_i\right) - H\left(Y_1, ..., Y_k\right)  + \delta\bigg)\bigg]
\nonumber
\\&= 
\exp\bigg[n\bigg(D\bigg(\bPP{Y_1, ..., Y_k} \| \prod_{i=1}^k\bPP{Y_i} \bigg) + \delta\bigg)\bigg],
\nonumber
\end{align}
for all $n$ large enough. Thus, the hypothesis of Theorem \ref{t:general_converse} holds with 
\begin{align}
\theta = \theta_n = \exp\bigg[n\bigg(D\bigg(\bPP{Y_1, ..., Y_k} \| \prod_{i=1}^k\bPP{Y_i} \bigg) + \delta\bigg)\bigg].
\nonumber
\end{align}
If $E$ is an $\ep$-achievable query exponent (see Definition \ref{d:query_exponent}),
then there exists an $\ep$-CR $L = L\left(Y_1^n, ..., Y_k^n\right)$ from communication $\mathbf{F} = \mathbf{F}\left(Y_1^n, ..., Y_k^n\right)$ such that (\ref{e:query_exp}) holds for the query strategy $q_0$ of Theorem \ref{t:general_converse} for this choice of $L$ and $\mathbf{F}$. In particular for $\ep^\prime < (1 - \delta	- \sqrt{\delta + \ep})^2$, we get from (\ref{e:q_adversary}) and (\ref{e:query_exp}) that 
\begin{align}
&\mathrm{P}\Bigg(\exp (nE) \leq q_0(L\mid \mathbf{F} ) \leq  
\delta^{-2/ (k-1)} \times
\nonumber
\\
& \qquad 
\exp \left[n\left(\frac{1}{k-1}D\bigg(\bPP{Y_1, ..., Y_k} \| \prod_{i=1}^k\bPP{Y_i} \bigg) + \frac{\delta}{k-1}\right)\right]\Bigg)  
\nonumber
\\&\geq (1 - \delta	- \sqrt{\delta + \ep})^2 - \ep^\prime
>0,
\label{e:32a}
\end{align}
for all $n$ sufficiently large. It follows that 
\begin{align}
E\leq \frac{1}{k-1}D\bigg(\bPP{Y_1, ..., Y_k} \| \prod_{i=1}^k\bPP{Y_i} \bigg) + \frac{2\delta}{k-1}.
\nonumber
\end{align}
Since $E$ was any $\ep$-achievable query exponent and $\delta>0$
was chosen arbitrarily, the assertion of Theorem \ref{t:converse} is established.\qed

{\it Proof of Theorem \ref{t:general_converse}.} Denote by $\cL$ the set of values of
the CR $L$. Using the hypothesis (\ref{e:Entropy_boundsOnP}) of the Theorem, we shall 
show below the existence of a set $\cI_o$ of values of $\bF$ and associated sets 
$\cL(\bi) \subseteq \cL$, $\bi \in \cI_0$, such that for every $\bi \in \cI_0$
\begin{align}
\bP{L \mid \bF}{\cL(\bi) \mid \bi} &\geq 1 - \delta - \sqrt{\ep + \delta},
\label{e:L(i)_lb_P}
\\|\cL(\bi)| &\leq \left(\frac{\theta}{\delta^2}\right)^{\frac{1}{k-1}},
\label{e:L(i)_ub_Card}
\\\text{and }\quad \bP{\bF}{\cI_0} &\geq 1 - \delta - \sqrt{\ep + \delta}.
\label{e:I0_lb_P}
\end{align}
Then, we consider a query strategy $q_0$ for $L$ given $\bF$ as in the
proof of converse part of Lemma \ref{l:equivalence}, with $L,\, \bF,\, \cI_0,\, \cL(\bi)$
in the roles of $U,\, V,\, \cV_0,\, \cU_v$, respectively. Thus, for all $\bi \in \cI_0$,
$l \in \cL(\bi)$,
\begin{align}
q_0(l\mid \bi) \leq |\cL(\bi)| \leq \left(\frac{\theta}{\delta^2}\right)^{\frac{1}{k-1}},
\nonumber
\end{align}
and so, as in (\ref{e:q_0_ub}), we get by (\ref{e:L(i)_lb_P})-(\ref{e:I0_lb_P}),
\begin{align}
\bPr{q_0(L\mid \mathbf{F} ) \leq  \left(\frac{\theta}{\delta^2}\right)^{\frac{1}{k-1}}}  \geq (1 - \delta	- \sqrt{\delta + \ep})^2,
\nonumber
\end{align}
thereby establishing the assertion (\ref{e:q_adversary}). 

The existence of the sets $\cI_0$ and $\{\cL(\bi), \bi\in \cI_0\}$ satisfying (\ref{e:L(i)_lb_P})-(\ref{e:I0_lb_P}) is argued
in three steps below.

\noindent{\it Step 1.} First, we note the following simple property of interactive communication: if rvs $Y_1, ..., Y_k$ are mutually independent,
they remain mutually independent when conditioned on an interactive communication $\bF$. 
\begin{lemma}\label{l:indep_condF}
Let the pmf $\tPP{Y_1, ..., Y_k}$ be such that 
\begin{align}
\tPP{Y_1, ..., Y_k} = \prod_{j=1}^k \tPP{Y_j}.
\label{e:Pt_def}
\end{align}
Then, for $\bi = \bF\left(y_1, ..., y_k\right)$, we have 
\begin{align}
\tP{Y_1, ..., Y_k\mid \bF}{y_1, ..., y_k\mid \bi} = \prod_{j=1}^k\tP{Y_j\mid \bF}{y_j\mid \bi}.
\label{e:Pt_condI}
\end{align}
\end{lemma}
{\it Proof.} The proof follows upon observing that
\begin{align}
&I_{\tPP{}}\left(Y_j\wedge Y_1, ..., Y_{j-1}, Y_{j+1}, ..., Y_k\mid \bF\right)
\nonumber
\\& \leq I_{\tPP{}}\left(Y_j\wedge Y_1, ..., Y_{j-1}, Y_{j+1}, ..., Y_k\right) 
\nonumber
\\&= 0, \quad j= 1, ..., k,
\label{e:independence_condF}
\end{align}
where the first inequality is by \cite[Lemma 2.2]{AhlCsi93} upon choosing $U = Y_j$, 
$V=\left( Y_1, ..., Y_{j-1}, Y_{j+1}, ..., Y_k\right)$,
$\Phi$ to be the communication from terminal $j$, and $\Psi$ to be the communication from the remaining terminals.

Hereafter in this proof, we shall select 
\begin{align}
\tPP{Y_j} = \bPP{Y_j}, \quad j =1, ..., k.
\label{e:Pt_marginals}
\end{align}

\noindent {\it Step 2.} In this step, we select the aforementioned set of communication values $\cI_0$. Let $L_j = L_j\left(Y_j, \bF\right)$
denote an estimate of CR $L$ at terminal $j$, $j =1, ..., k$ (see Definition \ref{d:CR_def}). 
Denote by $\cT_0$ the set $\{\cdot\}$ on the left side of (\ref{e:Entropy_boundsOnP}).
For each realization $(l, \bi)$ of $(L, \bF)$, denote by $A_{l, \bi} \subseteq \cY_1\times...\times \cY_k$ the set
\begin{align}
A_{l, \bi} = &\cT_0\cap \left\{ \left(y_1, ..., y_k\right): \bF\left(y_1, ..., y_k\right) = \bi, \right.
\nonumber
\\
&\hspace{0.5cm} \left.L_j\left(y_j, \bi\right)  = L\left(y_1, ..., y_k\right) = l, j = 1, ..., k \right\}.
\label{e:A_sets_definition}
\end{align}
Since $L$ is $\ep$-CR from $\bF$, we have from (\ref{e:CR_recover}) and (\ref{e:Entropy_boundsOnP}) that
\begin{align}
\bPr{\left(Y_1, ..., Y_k\right)\in A_{L, \bF}} \geq 1 - \ep - \delta.
\nonumber
\end{align}
By a reverse Markov inequality, there exists a set $\cI_1$ of values of $\bF$ with 
\begin{align}
\bP{\bF}{\cI_1} \geq 1 - \sqrt{\ep + \delta},
\label{e:I1_lb_P}
\end{align}
and 
\begin{align}
\bPr{\left(Y_1, ..., Y_k\right)\in A_{L, \bF} \mid \bF = \bi} \geq 1 - \sqrt{\ep + \delta},\,\, \bi \in \cI_1.
\label{e:I1_lb_condP}
\end{align}
Next, denote by $\cI_2$ the set of values of $\bF$ such that 
\begin{align}
\delta\tP{\bF}{\bi} \leq \bP{\bF}{\bi}, \qquad \bi \in \cI_2,
\label{e:I2_def}
\end{align}
where $\tPP{\bF}$ is, as usual, the distribution of $\bF$ under $\tPP{}$.
From Proposition \ref{p:measure_change} with $Q_1 = \bPP{\bF}$,
$Q_2 = \tPP{\bF}$, we have 
\begin{align}
\bP{\bF}{\cI_2} \geq 1 - \delta.
\label{e:I2_lb_P}
\end{align}
Thus, by (\ref{e:I1_lb_P}) and (\ref{e:I2_lb_P}), $\cI_0 \triangleq \cI_1 \cap \cI_2$ satisfies (\ref{e:I0_lb_P}).

\noindent {\it Step 3.} In this step, we identify sets $\cL(\bi)$ that satisfy (\ref{e:L(i)_lb_P}) and (\ref{e:L(i)_ub_Card}). For
each $\bi \in \cI_0$, the sets $A_{l, \bi}$ corresponding to different values $l$ are disjoint. Upon defining the 
nonnegative measure\footnote{Although $\mu$ depends on $\bi$, our notation will suppress this dependence.} $\mu$ on $\cL$ for each $\bi \in \cI_0$ by 
\begin{align}
\mu(l) \triangleq \bP{Y_1, ..., Y_k\mid \bF}{A_{l, \bi}\mid \bi}, \qquad l\in \cL,
\label{e:mu_def1}
\end{align}
we get 
\begin{align}
\mu(\cL) &= \sum_{l \in \cL}\bP{Y_1, ..., Y_k \mid \bF}{A_{l, \bi}\mid \bi} 
\nonumber
\\&= \bPr{\left(Y_1, ..., Y_k\right) \in A_{L,\bi} \mid \bF = \bi}
\nonumber
\\&\geq 1 - \sqrt{\ep + \delta},
\nonumber
\end{align}
by (\ref{e:I1_lb_condP}).
Applying Lemma \ref{l:card_largePset} (i) with $\cL$ in the role of $\cU$, we set
$\cL(\bi) = \cU_\delta$, and so
\begin{align}
\mu(\cL(\bi)) &\geq \mu(\cL) - \delta
\nonumber
\\&\geq 1 - \delta - \sqrt{\ep + \delta}
\label{e:L(i)_lb_P_proof}
\end{align}
and 
\begin{align}
|\cL(\bi)| \leq \delta^{-\alpha/  (1-\alpha)} \exp\left(H_\alpha(\mu)\right), \quad 0\leq \alpha <1.
\label{e:L(i)_ub_Renyi}
\end{align}
It follows from (\ref{e:L(i)_lb_P_proof}) that
\begin{align}
\bP{L\mid \bF}{\cL(\bi) \mid \bi} &\geq \sum_{l \in \cL(\bi)}\bP{Y_1, ..., Y_k\mid \bF}{A_{l, \bi}\mid \bi} 
\nonumber
\\&= \mu(\cL(\bi))
\nonumber
\\&\geq 1 - \delta - \sqrt{\ep + \delta},
\label{e:lb_P_LgivenF}
\end{align}
which establishes (\ref{e:L(i)_lb_P}).

Finally, we obtain an upper bound on $\exp\left(H_\alpha(\mu)\right)$ for $\alpha = \frac{1}{k}$, 
which will lead to (\ref{e:L(i)_ub_Card}). Denote by $A_{l, \bi}^j \subseteq \cY_j$ the projection of
the set $A_{l, \bi} \subseteq \cY_1 \times ... \times \cY_k$ along the $j$th coordinate, $j =1, ..., k$.
The sets $A_{l, \bi}^j$ are disjoint for different values of $l$, by definition (see (\ref{e:A_sets_definition})).
Thus, for the pmf $\tPP{Y_1, .., Y_k}$ in (\ref{e:Pt_def}), (\ref{e:Pt_marginals}), we have
\begin{align}
1 &\geq \prod_{j=1}^k \left[\sum_{l \in \cL} \tP{Y_j\mid \bF}{A_{l, \bi}^j \mid \bi}\right]
\nonumber
\\&\geq \left[\sum_{l \in \cL} \left(\prod_{j=1}^k \tP{Y_j\mid \bF}{A_{l, \bi}^j \mid \bi}^{\frac{1}{k}}\right)\right]^k, 
\label{e:Holders}
\end{align}
where the last step follows from H\"older's inequality\footnote{See \cite[equation (33)]{VenAna98} for an early 
use of H\"older's inequality in a CR converse proof.} \cite[Section 2.7]{HarLitPol52}. Using (\ref{e:Pt_condI}),
the right-side of (\ref{e:Holders}) is the same as
\begin{align}
\left[\sum_{l \in \cL} \tP{Y_1, ..., Y_k\mid \bF}{A_{l, \bi}^1 \times ... \times A_{l, \bi}^k \mid \bi}^{\frac{1}{k}}\right]^k, 
\nonumber
\end{align}
which is bounded below by
\begin{align}
\left[\sum_{l \in \cL}  \tP{Y_1, ..., Y_k\mid \bF}{A_{l, \bi} \mid \bi}^{\frac{1}{k}}\right]^k, 
\label{e:lb_condP_givenF_A}
\end{align}
since 
\begin{align}
A_{l, \bi} \subseteq A_{l, \bi}^1 \times ... \times A_{l, \bi}^k. 
\label{e:A_projection_subset}
\end{align}
Upon noting that $A_{l, \bi}\subseteq \cT_0$, for all $\left(y_1, ..., y_k\right) \in A_{l, \bi}$, it follows  that
\begin{align}
\tP{Y_1, ..., Y_k\mid \bF}{y_1, ..., y_k \mid \bi} &= \frac{\tP{Y_1, ..., Y_k}{y_1, ..., y_k }}{\tP{\bF}{\bi}}
\nonumber
\\&= \frac{\prod_{j=1}^k\tP{Y_j}{y_j}}{\tP{\bF}{\bi}}
\nonumber
\\&= \frac{\prod_{j=1}^k\bP{Y_j}{y_j}}{\tP{\bF}{\bi}}
\nonumber
\\&\geq \frac{\bP{Y_1, ..., Y_k}{y_1, ..., y_k }}{\theta\,\,\tP{\bF}{\bi}} 
\nonumber
\\&\geq \frac{\bP{Y_1, ..., Y_k\mid \bF}{y_1, ..., y_k \mid \bi}}{\delta^{-1}\,\,\theta},
\nonumber
\end{align}
where the third equality and the subsequent inequalities are by
(\ref{e:Pt_marginals}), (\ref{e:Entropy_boundsOnP}) and (\ref{e:I2_def}), respectively.
Combining the observations above with (\ref{e:Holders}) and (\ref{e:lb_condP_givenF_A}), 
we get
\begin{align}
1 &\geq \left[\sum_{l \in \cL}  \left(\frac{\bP{Y_1, ..., Y_k\mid \bF}{A_{l, \bi} \mid \bi}}{\delta^{-1}\,\,\theta}\right)^{\frac{1}{k}}\right]^k, 
\nonumber
\\&= \frac{\delta}{\theta}\left[\sum_{l \in \cL}\mu(l)^\frac{1}{k}\right]^k,
\nonumber
\end{align}
which, recalling Definition \ref{d: Renyi_Entropy}, further yields
\begin{align}
\exp\left(H_\frac{1}{k}(\mu)\right) &= \left[\sum_{l \in \cL}\mu(l)^\frac{1}{k}\right]^{\frac{k}{k -1}}
\nonumber
\\&\leq \left(\frac{\theta}{\delta}\right)^{\frac{1}{k-1}}.
\nonumber
\end{align}
The previous bound, along with (\ref{e:L(i)_ub_Renyi}), gives (\ref{e:L(i)_ub_Card}). \qed


\section{Converse Proof of Theorem \ref{t:main} for arbitrary $\cA\subseteq \cM$}\label{s:converse_A<M}
The converse technique of the previous section for $\cA = \cM$ can be extended 
to an arbitrary $\cA \subseteq \cM$, yielding an analogous upper 
bound for $E^*(\ep)$ in terms of divergences. However, the resulting
upper bound is inadequate as it is known to exceed the expression 
in the right-side of (\ref{e:E*_alt1}) (see \cite{Cha10}). In  this section,
we develop a new converse technique that targets directly the latter.

The main steps of the general converse proof for the case 
$\cA \subseteq \cM$ are analogous to those in the previous 
section. The central step is the counterpart of Theorem \ref{t:general_converse},
which is given next.
Given a fractional partition $\lambda$ as in (\ref{e:Lambda(A)}),
its dual partition is $\bl = \bl(\la) = \left\{\bl_{B^c}, B \in \cB\right\}$ with
\begin{align}
\bl_{B^c} = \frac{\lambda_B}{\ls - 1}, \quad B \in \cB,
\label{e:lambda_dual_def}
\end{align}
where $\cB$ is defined in (\ref{e:set_B}) and $\ls$ is given by (\ref{e:lambda_sum}). 
It is known from \cite{MadTet10},
and can be seen also from (\ref{e:Lambda(A)}) and (\ref{e:lambda_sum}),
that
\begin{align}\nonumber
\sum_{B\in \cB: B^c\ni i}\bl_{B^c} &=  \frac{1}{\ls -1}\sum_{B\in \cB: B^c\ni i}\lambda_B\\\nonumber
&=  \frac{1}{\ls -1}\left[\sum_{B\in \cB}\lambda_B - \sum_{B\in \cB: B\ni i}\lambda_B\right]\\
&=  \frac{1}{\ls -1} \left[\ls -1\right]
=1, \quad i \in \cM,
\label{e:dual_cond}
\end{align}
so that $\bl$, too, is a fractional partition of $\cM$.

\begin{theorem}\label{t:general_converse2}
Let $L =  L \left(Y_1, ..., Y_m\right)$ be $\ep$-CR
for $\cA$ from interactive communication 
$\bF = \bF\left(Y_1, ..., Y_m\right)$,
$0 < \ep < 1$.
 Given $\delta> 0$ such that $\delta + \sqrt{\delta + \ep} <1$
and a fractional partition $\lambda\in \Lambda(\cA)$, 
let $\theta_{B^c}, B \in \cB$, and $\theta_0$ be such that 
\begin{align}
&\bPr{\left\{ y_\cM: \bP{Y_\cM}{y_\cM} \leq \frac{1}{\theta_0}, \,\,
\bP{Y_{B^c}}{y_{B^c}}  \geq \frac{1}{\theta_{B^c}}, B \in \cB \right\}}
\nonumber
\\&\geq 1 - \delta.
\label{e:Entropy_boundsOnP2}
\end{align}
Then, with
\begin{align}
\theta = \frac{\displaystyle \prod_{B \in \cB}\,\, \theta_{B^c}^{\bl_{B^c}}}{\theta_0},
\label{e:theta_def}
\end{align}
there exists a query strategy $q_0$ for $L$ given $\mathbf{F}$
such that
\begin{align}
\bPr{q_0(L\mid \mathbf{F} ) \leq  \left(\frac{\theta}{\kappa(\delta)}\right)^{\ls -1}}  \geq (1 - \delta	- \sqrt{\delta + \ep})^2,
\label{e:q_adversary2}
\end{align}
where $\kappa(\delta) = \left(m2^m\right)^{-m}\delta^{m+1}$.
\end{theorem}
{\it Proof.} As in the proof of Theorem \ref{t:general_converse}, the assertion (\ref{e:q_adversary2}) will follow upon showing the existence of sets $\cI_0$
and $\cL(\bi) \subseteq \cL$, $\bi \in \cI_0$, such that (\ref{e:L(i)_lb_P}) and 
(\ref{e:I0_lb_P}) are satisfied, along with the following replacement for (\ref{e:L(i)_ub_Card}):
\begin{align}
|\cL(\bi)| \leq \left(\frac{\theta}{\kappa(\delta)}\right)^{\ls - 1}, \qquad \bi \in \cI_0.
\label{e:L(i)_ub_Card2}
\end{align} 
To this end, we provide here appropriate replacements for the three steps in the proof of Theorem \ref{t:general_converse}. 

\noindent {\it Step 1.} For each $B \subsetneq \cM$, consider the pmf $\tPP{Y_\cM}^B$ defined by 
\begin{align}
\tPs{B}{Y_\cM}{y_\cM} = \bP{Y_B}{y_B}\bP{Y_{B^c}}{y_{B^c}}
\label{e:Pt_def2}
\end{align}
Note that $\tPP{}^B \equiv \tPP{}^{B^c}$. The collection 
of pmfs $\left\{ \tPP{}^{B^c},\,\, B \in \cB \right\}$ serve as a replacement for the pmf $\tPP{}$
in (\ref{e:Pt_def}). 

For the pmf $\tPP{}^B$ in (\ref{e:Pt_def2}), we note that 
\begin{align}
I_{\tPP{}^B}\left(Y_B \wedge F_{kj}\mid \Phi_{kj}\right) = 0, \qquad  j \in B^c,
\label{e:independence_condF2}
\end{align}
since $F_{kj} = f_{kj}\left(Y_j, \Phi_{kj}\right)$ and $Y_{B^c}$ is independent
of $Y_B$ conditioned on $\Phi_{kj}$.
The following Lemma serves 
the role of Lemma \ref{l:indep_condF}.

\begin{lemma}\label{l:indep_condF2}
For $B \subsetneq \cM$ and $\bi = \bF\left(y_\cM\right)$, we have 
\begin{align}\ds
\tPs{B}{Y_B\mid \bF}{y_B \mid \bi} = 
\frac{\bP{Y_B}{y_B}}{\prod_{k=1}^r \prod_{j \in B}\tPs{B}{F_{kj}\mid \Phi_{kj}}{i_{kj}\mid i_{kj}^-}},
\label{e:independence_condF3}
\end{align}
where $i_{kj}^-$ denotes the past values of communication 
in $\bi$ for 
round $k$ and terminal $j$.
\end{lemma}
{\it Proof.} Note that 
\begin{align}\ds
\tPs{B}{Y_B\mid \bF}{y_B \mid \bi} &= 
\frac{\tPs{B}{\bF\mid Y_B}{\bi\mid y_B}\tPs{B}{Y_B}{y_B}}{\tPs{B}{\bF}{\bi}}
\nonumber\\
 &= 
\frac{\tPs{B}{\bF\mid Y_B}{\bi\mid y_B}\bP{Y_B}{y_B}}{\tPs{B}{\bF}{\bi}},
\label{e:l9_1}
\end{align}
where the previous step is by (\ref{e:Pt_def2}). Furthermore, 
\begin{align}\ds
\tPs{B}{\bF\mid Y_B}{\bi\mid y_B} &= 
\prod_{k=1}^r \prod_{j =1}^m\tPs{B}{F_{kj}\mid Y_B, \Phi_{kj}}{i_{kj}\mid y_B, i_{kj}^-}
\nonumber\\
 &= 
\prod_{k=1}^r \prod_{j \in B^c}\tPs{B}{F_{kj}\mid Y_B, \Phi_{kj}}{i_{kj}\mid y_B, i_{kj}^-}
\nonumber\\
 &= 
\prod_{k=1}^r \prod_{j \in B^c}\tPs{B}{F_{kj}\mid \Phi_{kj}}{i_{kj}\mid i_{kj}^-},
\label{e:l9_2}
\end{align}
where the last step uses (\ref{e:independence_condF2}). Next, 
\begin{align}\ds
&\tPs{B}{\bF}{\bi} 
\nonumber
\\&=\prod_{k=1}^r \prod_{j =1}^m\tPs{B}{F_{kj}\mid \Phi_{kj}}{i_{kj}\mid i_{kj}^-}
\nonumber\\
&= \prod_{k=1}^r \left(\prod_{j \in B}\tPs{B}{F_{kj}\mid \Phi_{kj}}{i_{kj}\mid i_{kj}^-}\prod_{j \in B^c}\tPs{B}{F_{kj}\mid \Phi_{kj}}{i_{kj}\mid i_{kj}^-}\right).
\label{e:l9_3}
\end{align}
Then (\ref{e:l9_1}), along with (\ref{e:l9_2}) and (\ref{e:l9_3}), gives (\ref{e:independence_condF3}). 

\noindent {\it Step 2.} Denoting by $\cT_0$ the set $\{\cdot\}$ on the left-side of (\ref{e:Entropy_boundsOnP2}), for each $L = l, \bF = \bi$, define 
\begin{align}
A_{l,\bi} =& \cT_0 \cap \left\{y_\cM: \bF\left(y_\cM\right)=\bi, \right.
\nonumber
\\
&\hspace*{1.6cm} \left.L_j\left(y_j, \bi\right) = L\left(y_\cM\right) = l, j \in \cA \right\}.
\label{e:A_sets_definition2}
\end{align}
Analogous to the proof of Theorem \ref{t:general_converse}, the set $\cI_1$ of values of $\bF$ with
\begin{align}
\bPr{Y_\cM \in A_{L, \bF} \mid \bF =\bi} \geq 1 - \sqrt{\ep+\delta}, \qquad \bi \in \cI_1,
\nonumber
\end{align}
satisfies 
\begin{align}
\bP{\bF}{\cI_1} \geq 1 - \sqrt{\ep+\delta}.
\nonumber
\end{align}
For $j \in \cM$ and $B \subsetneq \cM$, denote by $\cI_{j, B}$ the set of 
$\bi$ such that
\begin{align}
&\left(m2^m\right)^{-1}\delta \prod_{k=1}^r \tPP{F_{kj}\mid \Phi_{kj}}^B\left(i_{kj}\mid i_{kj}^-\right) 
\nonumber
\\&\leq \prod_{k=1}^r \bP{F_{kj}\mid \Phi_{kj}}{i_{kj}\mid i_{kj}^-}.
\label{e:Pt_to_P}
\end{align}
The following simple extension of Proposition \ref{p:measure_change} holds:
\begin{align}
&\bP{\bF}{\cI_{j, B}^c} 
\nonumber\\
&= \sum_{\bi \in \cI_{j,B}^c}\bP{\bF}{\bi}
\nonumber\\
&= \sum_{\bi \in \cI_{j,B}^c}\prod_{l=1}^m\prod_{k=1}^r\bP{F_{kl}\mid \Phi_{kl}}{i_{kl}\mid i_{kl}^-}
\nonumber\\
&= \sum_{\bi \in \cI_{j,B}^c}\left(\prod_{l\neq j}\prod_{k=1}^r\bP{F_{kl}\mid \Phi_{kl}}{i_{kl}\mid i_{kl}^-}\right)\times
\nonumber\\
&\hspace*{3cm}\prod_{k=1}^r\bP{F_{kj}\mid \Phi_{kj}}{i_{kj}\mid i_{kj}^-}
\nonumber\\
&< \left(m2^m\right)^{-1}\delta\sum_{\bi \in \cI_{j,B}^c}\left(\prod_{l\neq j}\prod_{k=1}^r\bP{F_{kl}\mid \Phi_{kl}}{i_{kl}\mid i_{kl}^-}\right) \times
\nonumber\\
&\hspace*{3cm}\prod_{k=1}^r \tPP{F_{kj}\mid \Phi_{kj}}^B\left(i_{kj}\mid i_{kj}^-\right)
\nonumber\\
&\leq \left(m2^m\right)^{-1}\delta\sum_{\bi}\left(\prod_{l\neq j}\prod_{k=1}^r\bP{F_{kl}\mid \Phi_{kl}}{i_{kl}\mid i_{kl}^-}\right) \times
\nonumber\\
&\hspace*{3cm}\prod_{k=1}^r \tPP{F_{kj}\mid \Phi_{kj}}^B\left(i_{kj}\mid i_{kj}^-\right)
\nonumber\\
&= \left(m2^m\right)^{-1}\delta,
\label{e:new_P=PtP}
\end{align}
where the first inequality is by (\ref{e:Pt_to_P}), and (\ref{e:new_P=PtP}) holds since the
summand is a pmf for $\bF$, as can be seen by directly computing the sum.
Defining $\ds\cI_2 = \bigcap_{j=1}^m \bigcap_{B \subsetneq \cM}\cI_{j, B}$, we get 
\begin{align}
\bP{\bF}{\cI_2} \geq 1 - \delta.
\nonumber
\end{align}
The set $\cI_0$ is defined as $\cI_1\cap \cI_2$, and satisfies (\ref{e:I0_lb_P}).

\noindent {\it Step 3.} Finally, we define sets $\cL(\bi)\subseteq \cL$, $\bi \in \cI_0$ that
satisfy (\ref{e:L(i)_lb_P}) and (\ref{e:L(i)_ub_Card2}). For each $\bi \in \cI_0$, let 
\begin{align}
\mu(l) = \bP{Y_\cM\mid \bF}{A_{l, \bi}\mid \bi}, \qquad l \in \cL.
\end{align}
Then, the sets $\cL(\bi)$ satisfying (\ref{e:L(i)_lb_P}) are obtained by an 
application of Lemma \ref{l:card_largePset} (i) as in (\ref{e:L(i)_lb_P_proof})
and (\ref{e:L(i)_ub_Renyi}) above.

The condition (\ref{e:L(i)_ub_Card2}) will be obtained upon showing that for 
\begin{align}
\alpha = \frac{\ls - 1}{\ls},
\label{e:alpha_def}
\end{align}
it holds that
\begin{align}
\delta^{-\alpha/  (1-\alpha)}\exp\left(H_\alpha(\mu)\right) \leq 
\left(\frac{\theta}{\kappa(\delta)}\right)^{\ls -1}.
\label{e:L(i)_ub_Renyi2}
\end{align}
To do so, first note that for each $B \in \cB$, the set $B^c \cap \cA$ is 
nonempty. Thus, by (\ref{e:A_sets_definition2}), the projections $A_{l, \bi}^{B^c}$ of $A_{l, \bi}$
along the coordinates in $B^c \subsetneq \cM$ are disjoint across $l \in \cL$.
Thus,
\begin{align}\ds
1 \geq \prod_{B \in \cB} \left(\sum_{l \in \cL}\tPs{B^c}{Y_\cM \mid \bF}{A_{l, \bi}^{B^c}\mid \bi}\right)^{\la_B}.
\nonumber
\end{align}
Using H\"older's inequality \cite[Section 2.7]{HarLitPol52}, and recalling (\ref{e:lambda_dual_def}) and (\ref{e:lambda_sum}) we get
\begin{align}\ds
1 \geq \left[\sum_{l \in \cL}\left(\prod_{B \in \cB} \tPs{B^c}{Y_\cM \mid \bF}{A_{l, \bi}^{B^c}\mid \bi}^{\bl_{B^c}}\right)^\alpha\,\right]^{\frac{1}{1-\alpha}}.
\label{e:B1}
\end{align}
Next, note from Lemma \ref{l:indep_condF2} that
\begin{align}
\tPs{B^c}{Y_\cM \mid \bF}{A_{l, \bi}^{B^c}\mid \bi} =
 \frac{\ds\sum_{y_{B^c}\in A_{l,\bi}^{B^c}}\bP{Y_{B^c}}{y_{B^c}}}{\ds\prod_{k=1}^r \prod_{j \in B^c}\tPs{B^c}{F_{kj}\mid \Phi_{kj}}{i_{kj}\mid i_{kj}^-}},
\nonumber
\end{align}
which, since the order of products can be interchanged, and upon using (\ref{e:Pt_to_P}), is bounded below by
\begin{align}\ds
 \frac{\ds\sum_{y_{B^c}\in A_{l,\bi}^{B^c}}\bP{Y_{B^c}}{y_{B^c}}}{\ds \prod_{j \in B^c}\left(m2^m\right) \delta^{-1}\,\ds \prod_{k=1}^r\left[\bP{F_{kj}\mid \Phi_{kj}}{i_{kj}\mid i_{kj}^-}\right]}.
\nonumber
\end{align}
It follows that
\begin{align}
& \prod_{B \in \cB} \tPs{B^c}{Y_\cM \mid \bF}{A_{l, \bi}^{B^c}\mid \bi}^{\bl_{B^c}}
\nonumber\\
& \geq 
 \frac{\ds
\prod_{B \in \cB} \left[\sum_{y_{B^c}\in A_{l,\bi}^{B^c}}\bP{Y_{B^c}}{y_{B^c}}
\,\right]^{\bl_{B^c}}}{\ds\prod_{B \in \cB}\prod_{j \in B^c}
\left[\left(m2^m\right) \delta^{-1}\,\ds\prod_{k=1}^r \bP{F_{kj}\mid \Phi_{kj}}{i_{kj}\mid i_{kj}^-}\right]^{\bl_{B^c}}}.
\label{e:B2}
\end{align}
The right-side of (\ref{e:B2}) can be simplified by noting that 
\begin{align}
&\prod_{B \in \cB} \prod_{j \in B^c}
\left[\left(m2^m\right) \delta^{-1}\,\prod_{k=1}^r\bP{F_{kj}\mid \Phi_{kj}}{i_{kj}\mid i_{kj}^-}\right]^{\bl_{B^c}}
\nonumber\\
&=\prod_{j =1}^m
\left[\left(m2^m\right) \delta^{-1}\,\prod_{k=1}^r \bP{F_{kj}\mid \Phi_{kj}}{i_{kj}\mid i_{kj}^-}\right]^{\sum_{B\in \cB: B^c\ni j}\bl_{B^c}}
\nonumber\\
&=\left(\frac{m2^m}{\delta}\right)^{m}\bP{\bF}{\bi},
\label{e:B3}
\end{align}
where the previous step uses (\ref{e:dual_cond}). The definition of $\cT_0$,
along with (\ref{e:B2}) and (\ref{e:B3}), gives 
\begin{align}
& \prod_{B \in \cB} \tPs{B^c}{Y_\cM \mid \bF}{A_{l, \bi}^{B^c}\mid \bi}^{\bl_{B^c}}
 \nonumber\\
 &\geq  \frac{\delta^{m}}{\left(m2^m\right)^{m}\bP{\bF}{\bi}}
 \prod_{B \in \cB} \left(\frac{ \big|A_{l,\bi}^{B^c}\big|}{\theta_{B^c}}\right)^{\bl_{B^c}}.
\label{e:B4}
\end{align}
Also, since $A_{l,\bi} \subseteq \cT_0$, we have 
\begin{align}
\bP{Y_\cM}{A_{l,\bi}} \leq \frac{\big|A_{l,\bi}\big|}{\theta_0},
\nonumber
\end{align}
which, with (\ref{e:theta_def}) and (\ref{e:B4}), gives 
\begin{align}
&\prod_{B \in \cB} \tPs{B^c}{Y_\cM \mid \bF}{A_{l, \bi}^{B^c}\mid \bi}^{\bl_{B^c}}
\nonumber\\
 &\geq  \frac{\delta^{m}}{\left(m2^m\right)^{m}\theta}
\left(\frac{  \prod_{B \in \cB} \big|A_{l,\bi}^{B^c}\big|^{\bl_{B^c}}}{\big|A_{l,\bi}\big|}\right)
\bP{Y_{\cM}\mid \bF}{A_{l,\bi}\mid \bi}.
\label{e:B5}
\end{align}
Since $\bl$ is a fractional partition, \cite[Corollary 3.4]{MadMarTet12} implies 
\begin{align}
\left(\frac{  \prod_{B \in \cB} \big|A_{l,\bi}^{B^c}\big|^{\bl_{B^c}}}{\big|A_{l,\bi}\big|}\right)\geq 1,
\label{e:B6}
\end{align}
which combined with (\ref{e:B1})-(\ref{e:B6}) yields
\begin{align}
1 \geq \left(\frac{\delta^{m}}{\left(m2^m\right)^{m}\theta}\right)^{\frac{\alpha}{1-\alpha}}
\left[\sum_{l\in \cL} \mu(l)^\alpha\,\right]^{\frac{1}{1-\alpha}}.
\nonumber
\end{align}
The previous inequality implies (\ref{e:L(i)_ub_Renyi2}) since
$$\frac{\alpha}{1-\alpha} = \ls -1.$$\qed


\section{Strong converse for secret key capacity}\label{s:strong_converse}

A byproduct of Theorem \ref{t:main} is a new result that establishes a strong 
converse for the SK capacity of a multiterminal source model, for the terminals in $\cA \subseteq \cM$. 
In this context, we 
shall consider -- without loss of effect -- a weaker notion of security index than in 
(\ref{e:s_in_def}), defined in terms of variational distance:
\begin{align}
s_{var}(K; \mathbf{F}) = \sum_{\bi} \bP{\bF}{\bi} \sum_{k=1}^{\|K\|} \left| \bP{K\mid \bF}{k\mid \bi} - \frac{1}{\|K\|}\right|.
\label{e:s_var_def}
\end{align}
However, the requirement (\ref{e:SK_security}) on $s_{in}$ will be 
replaced now by
\begin{align}
\lim_{n} n s_{var}(K; \bF) = 0.
\label{e:SK_secruity_new}
\end{align}
\begin{definition}\label{d:SK_cap}
Given $0< \ep< 1$, $R\geq 0$ is an $\ep$-achievable SK rate for $\cA \subseteq \cM$ 
if for every $\rho > 0$, there is an $N = N(\ep, \rho)$ such that for every $n \geq N$, there 
exists an $\ep$-CR $K =K\left(X_\cM^n\right)$ for $\cA$ from $\bF$ satisfying
\begin{align}
\frac{1}{n}\log \|K\| \geq R- \rho,
\label{e:SK_rate}
\end{align}
and 
\begin{align}
s_{var}(K; \bF) \leq \frac{\rho}{n}.
\label{e:SK_security_rho}
\end{align}
\end{definition}
The supremum of $\ep$-achievable SK rates is the $\ep$-SK capacity, denoted $C(\ep)$. The 
SK capacity is the infimum of $C(\ep)$ for  $0< \ep <1$. We recall the following.

\begin{theorem}\cite{CsiNar04}\label{t:SK_capacity}
The secret key capacity for $\cA \subseteq \cM$ is 
\begin{align}
C = E^* = H\left(X_\cM\right) - &\max_{\la \in \Lambda(\cA)} 
\sum_{B\in \cB }\lambda_B 
H\left(X_B\mid X_{B^c}\right), 
\nonumber\\
&\hspace*{3cm} 0<\ep< 1.
\nonumber
\end{align}
\end{theorem}
\begin{remark*}
The (new) secrecy requirement (\ref{e:SK_secruity_new}) is not unduly
restrictive. Indeed, the achievability proof of Theorem \ref{t:SK_capacity} \cite{CsiNar04}
holds with $s_{in}(K; \bF)$ vanishing to zero exponentially rapidly in $n$, which, by
Pinsker's inequality (cf. \cite{CsiKor11}), implies (\ref{e:SK_secruity_new}). 
The converse proof in \cite{CsiNar04} was shown under the ``weak secrecy"
condition
\begin{align}
\lim_n \frac{1}{n} I(K \wedge \bF) = 0,
\label{e:weak_secrecy}
\end{align}
which, in turn, is implied by (\ref{e:SK_secruity_new}) by a simple
application of \cite[Lemma 1]{CsiNar04}.
\end{remark*}
The strong converse for SK capacity, valid under (\ref{e:SK_secruity_new}), is 
given next.
\begin{theorem} \label{t:strong_converse_SK_cap}
For every $0< \ep < 1$, it holds that 
\begin{align}
C(\ep) = C.
\label{e:SK_capacity_strong}
\end{align}
\end{theorem}
\begin{remark*}
It is not known if the strong converse in Theorem \ref{t:strong_converse_SK_cap} 
holds under (\ref{e:weak_secrecy}).
\end{remark*}
{\it Proof.} Theorem \ref{t:SK_capacity} \cite{CsiNar04} already provides the
proof of achievability, i.e., $C(\ep) \geq C$. The converse proof below shows that if $R$ is an $\ep$-achievable
SK rate, then $R$ is an $\ep$-achievable query exponent. Therefore,
\begin{align}
R\leq E^*(\ep) = C,\quad 0 < \ep < 1,
\label{e:ub_C(e)}
\end{align}
where the equality is by (\ref{e:E*_char}). Specifically, for every $\rho > 0$, suppose that
there exists $K =K\left(X_\cM^n\right)$ and communication $\bF$ satisfying (\ref{e:SK_rate}) and (\ref{e:SK_security_rho}) for all $n$ sufficiently large. We claim that the hypothesis (\ref{e:ub1_condP}) of Lemma \ref{l:equivalence} holds with $U=K$, $V=\bF$ and 
$\gamma = \exp[n(R - 2\rho)]$ for every $0< \delta < 1/ 2$, when $\rho$ is sufficiently small. 
Therefore, by (\ref{e:lb_query}), $R-2\rho$ is an $\ep$-achievable query exponent which leads to 
(\ref{e:ub_C(e)}) since $\rho$ can be chosen arbitrarily small.

Turning to the claim, observe that
\begin{align}
&\bPr{\left\{(k, \bi): \bP{K\mid \bF}{k\mid \bi} > \frac{2}{\exp[n(R-\rho)]}\right\}}
\nonumber\\
&\leq \bPr{\left\{(k, \bi): \bP{K\mid \bF}{k\mid \bi} > \frac{2}{\|K\|}\right\}}
\nonumber\\
&\leq \bPr{\left\{(k, \bi): \big|\log\|K\|\bP{K\mid \bF}{k\mid \bi}\big| > 1\right\}}
\nonumber\\
&\leq  \bEE{\big|\log\|K\|\bP{K\mid \bF}{K\mid \bF}\big|},
\nonumber
\end{align}
where the first and the last inequality above follow from (\ref{e:SK_rate}) and the Markov inequality,
respectively. 

Next, we show that 
\begin{align}
\bEE{\big|\log\|K\|\bP{K\mid \bF}{K\mid \bF}\big|} \leq 
s_{var}(K; \bF)\log \frac{\|K\|^2}{s_{var}(K; \bF)}.
\label{e:ub_Expectation}
\end{align}
Then, the right-side can be bounded above by 
\begin{align}
\frac{\rho}{n}\log\frac{n}{\rho} + 2\rho\log \big|X_\cM\big|,
\label{e:ub_card_K}
\end{align}
for all $n$ sufficiently large; the claim follows upon taking 
$n\rightarrow \infty$ and $\rho \rightarrow 0$. To see (\ref{e:ub_Expectation}), 
note that for $t_1, t_2,$ $|t_1 - t_2| < 1$, $f(t) \triangleq -t\log t$ satisfies 
(cf. \cite[Lemma 2.7]{CsiKor11})
\begin{align}
\big|f(t_1) - f(t_2)\big| \leq \big| t_1 - t_2\big| \log \frac{1}{\big| t_1 - t_2\big|}.
\label{e:observation_concave_f}
\end{align}
Then, for $\bF = \bi$,
\begin{align}
&\sum_k \bP{K\mid \bF}{k\mid \bi}
\left|\log\|K\|\bP{K\mid \bF}{k\mid \bi}\right|
\nonumber\\
&= \sum_k 
\left|\bP{K\mid \bF}{k\mid \bi}\log\bP{K\mid \bF}{k\mid \bi}
+ \bP{K\mid \bF}{k\mid \bi} \log \|K\|\right.
\nonumber\\
&\hspace*{4cm}\left.+ \frac{1}{\|K\|}\log \|K\| - \frac{1}{\|K\|}\log \|K\|
\right|
\nonumber\\
&\leq \sum_k \left[
\left|\bP{K\mid \bF}{k\mid \bi}\log\bP{K\mid \bF}{k\mid \bi} - \frac{1}{\|K\|}\log\frac{1}{\|K\|}\right| \right.
\nonumber\\
&\hspace*{3.8cm}\left.+ \left|\bP{K\mid \bF}{k\mid \bi} - \frac{1}{\|K\|}\right|\log\|K\|\right]
\nonumber \\
&\leq \sum_k 
 \left|\bP{K\mid \bF}{k\mid \bi} - \frac{1}{\|K\|}\right|\log\frac{\|K\|}{\left|\bP{K\mid \bF}{k\mid \bi} - \frac{1}{\|K\|}\right|},
\label{e:ub_Expectation1}
\end{align}
where the previous inequality uses (\ref{e:observation_concave_f}) with 
$t_1 = \bP{K\mid \bF}{k\mid \bi}$ 
and $t_2 = \|K\|^{-1}$ for every value $k$
of $K$. 
Finally, (\ref{e:ub_Expectation}) follows upon 
multiplying both sides by $\bP{\bF}{\bi}$, summing over 
$\bi$ and using the log-sum inequality \cite{CsiKor11}.\qed

Observe that the proof of Theorem \ref{t:strong_converse_SK_cap} does
not rely on the form of the rvs $K, \bF$, and is, in effect, a statement 
relating the {\it size} of any achievable SK rate under the $s_{var}$-secrecy requirement (\ref{e:SK_secruity_new}) to the query exponent. As a consequence, also
the SK capacity for more complex models in which the eavesdropper
has additional access to side information can be bounded above
by the optimum query exponent when the querier, too, is given access to the same
side information.

\section{General Alphabet Converse for $\cA =\cM$}\label{s:general_converse}

In this section, we present a converse technique for 
the optimum query exponent for rvs with general alphabets,
with jointly Gaussian rvs as a special case. No corresponding 
general claim is made regarding achievability of the exponent.
Our technique also leads to a new strong converse for Gaussian
SK capacity \cite{NitNar12}.

Let $\cY_i$ be a complete separable metric space, with associated Borel $\sigma$-field
$\sigma_i$, $1 \leq i \leq k$; a special case of interest is  $\cY_i = \mathbb{R}^{n_i}$.
Denote by $\cY^k$ the set $\cY_1\times ... \times \cY_k$ and
by $\sigma^k$ the product $\sigma$-field\footnote{Hereafter, the 
term ``product $\sigma$-field" of $\sigma$-fields $\sigma_1, ..., \sigma_k$,
will mean the smallest $\sigma$-field containing 
sets from $\sigma_1\times ... \times\sigma_k$, and will be denoted, with an abuse of notation, simply as 
$\sigma^k = \sigma_1 \times ... \times \sigma_k$.
}
$\sigma_1 \times ... \times \sigma_k$ on $\cY^k$. Let $\mathrm{P} = \bPP{Y_1, ..., Y_k}$ 
be a probability measure on $\left(\cY^k, \sigma^k\right)$.
The interactive communication
$\left\{F_{ji}: 1\leq j \leq r, 1\leq i\leq k \right\}$ is specified as in Definition \ref{d:PubComm},
with the rv $F_{ji}$ taking values in, say, $\left(\cZ_{ji}, \cF_{ji}\right)$, and being $\sigma_i$-measurable 
for each fixed value of the preceding communication 
$$\Phi_{ji} = \left(F_{st}: 1\leq s < j, 1\leq t\leq k \text{ or } s=j, 1\leq t < i\right).$$
Then, there exists a unique regular conditional probability measure 
on $\left(\cY^k, \sigma^k\right)$ conditioned on $\sigma(\bF)$, denoted 
$\bPP{Y_1, ..., Y_k\mid \bF}$ (cf. \cite[Chapter 6]{Ash72}). 
The notation $Q_\bi$ will be used interchangeably for the probability measure 
$\bP{Y_1, ..., Y_k\mid \bF}{\cdot \mid \bi}$.
We make the following basic assumption
of absolute continuity:
\begin{align}
Q_\bi << \bPP{Y_1, ..., Y_k}, \qquad \bPP{\bF} \text{ a.s. in } \bi,
\label{e:assumption_general}
\end{align}
i.e., (\ref{e:assumption_general}) holds over a set of $\bi$ with $\bPP{\bF}$-probability 1.
Assumption (\ref{e:assumption_general}) is satisfied by a large class of
interactive communication protocols including $\bF$ taking countably 
many values. Moreover, we can assume the following without loss of generality:
\begin{align}
Q_\bi\left(\bF^{-1}(\bi)^c \right) &= 0, \qquad \bPP{\bF}\text{ a.s. in } \bi,
\label{e:assumption_general2i}\\
\frac{d\, Q_\bi}{d\, \mathrm{P}}(y^k) &= 0,\quad \text{ for } y^k \in \bF^{-1}(\bi)^c,\,\, \bPP{\bF}\text{ a.s. in } \bi.
\label{e:assumption_general2ii}
\end{align}

Next, we define $\ep$-CR $L$ from $\bF$ and its local
estimates $L_i$, respectively, as rvs {\it taking countably many values}, measurable 
with respect to $\sigma^k$ and $\sigma_i\times \sigma(\bF)$, $1\leq i \leq k$, and
satisfying
\begin{align}
\bPr{L = L_i,\, 1\leq i \leq k} \geq 1 - \ep.
\nonumber
\end{align}
The main result of this section, given below, extends Theorem \ref{t:general_converse}
to general measures as above.
\begin{theorem}\label{t:very_general_converse}
For $0< \ep <1$, let $L$ be $\ep$-CR from interactive communication $\mathbf{F}$.
Let $\tilde{\mathrm{P}} = \tPP{Y_1, ..., Y_k}$ be a probability measure on $\left(\cY^k, \sigma^k\right)$ with 
\begin{align}
\tilde{\mathrm{P}}\left(A_1\times ... \times A_k\right) = \prod_{i=1}^k \bP{Y_i}{A_i} \qquad A_i \in \sigma_i, \,1\leq i\leq k.
\label{e:tP_def3}
\end{align}
Assuming that $\mathrm{P}<< \tilde{\mathrm{P}}$,
and given $\delta> 0$ such that $\delta + \sqrt{\delta + \ep} <1$, let $\theta$ be such that 
\begin{align}
\bPr{\left\{ y^k: \frac{d\,\mathrm{P}}{d\,\tilde{\mathrm{P}}}(y^k) \leq \theta\right\}} \geq 1 - \delta.
\label{e:Entropy_boundsOnP3}
\end{align}
Then, there exists a query strategy $q_0$ for $L$ given $\mathbf{F}$
such that
\begin{align}
\bPr{q_0(L\mid \mathbf{F} ) \leq  \left(\frac{\theta}{\delta^2}\right)^{\frac{1}{k-1}}}  \geq (1 - \delta	- \sqrt{\delta + \ep})^2.
\label{e:q_adversary3}
\end{align}
\end{theorem}
The proof of Theorem \ref{t:very_general_converse} is deferred to the 
end of this section. At this point, we present its implications for a 
Gaussian setup. Let $X_i^{(n)}$ be an $\mathbb{R}^n$-valued rv,
$i=1, ..., m$, and let $X_\cM^{(n)} = \left(X_1^{(n)}, ..., X_m^{(n)}\right)$ be jointly Gaussian 
$\cN(\mathbf{0}, \Sigma^{(n)})$, where $\Sigma^{(n)}$ is a positive definite 
matrix. We remark that $X_\cM^{(n)}$
need not be independent or identically distributed across $n$. The notion
of an $\ep$-optimum query exponent $E^*(\ep)$, $0 <  \ep <1$, is exactly as in Definition \ref{d:query_exponent}, even though the
underlying CR now can take countably many values. Also, given a partition 
$\pi$ of $\cM$ with $|\pi| = k$, $2\leq k \leq m$,
the quantity $E^*_\pi(\ep)$ is defined as in Section \ref{s:converse_A=M}.

\begin{proposition}\label{p:ub_E_Gaussian}
For $X_\cM^{(n)} \sim \cN(\mathbf{0}, \Sigma^{(n)})$ with $\Sigma^{(n)}$
being positive definite, it holds that
\begin{align}
&E^*(\ep)\leq \min_{\pi} E^*_\pi(\ep) 
\nonumber\\
&\leq \min_{\pi} \frac{1}{2(|\pi| - 1)} \limsup_n \frac{1}{n}
\log\frac{ \prod_{i=1}^{|\pi|} \big|\Sigma^{(n)}_{\pi_i}\big|}{|\Sigma^{(n)}	|},\quad
0 < \ep <1,
\nonumber
\end{align}
where $\Sigma^{(n)}_{\pi_i}$ is the covariance matrix of $X^{(n)}_{\pi_i}$, $i = 1, ..., |\pi|$, and $|\cdot|$
denotes determinant.
\end{proposition}
\begin{corollary}
When $X_\cM^{(n)}$ is i.i.d. in $n$ with $X_\cM \sim \cN(\mathbf{0}, \Sigma)$,
\begin{align}
E^*(\ep) \leq \min_\pi \frac{1}{2(|\pi| -1)} \log\frac{\prod_{i=1}^{|\pi|}
\big|\Sigma_{\pi_i}\big|}{|\Sigma|}, \quad 0 < \ep <1.
\nonumber
\end{align}
\end{corollary}
{\it Proof.} Proceeding as in the proof of 
Theorem \ref{t:converse}, we apply Theorem \ref{t:very_general_converse} to 
the rvs $Y_i = X_{\pi_i}^{(n)}$, $1 \leq i \leq |\pi|$. Specifically, we show that the hypothesis 
(\ref{e:Entropy_boundsOnP3}) is satisfied with 
\begin{align}
\theta = \theta_n = 
\left(\frac{\prod_{i=1}^{|\pi|} \big|\Sigma_{\pi_i}^n\big|}{|\Sigma^{(n)}|}\right)^{1/ 2}\,\,\exp (n\delta),
\label{e:theta_n_Gaussian}
\end{align}
where $0 < \delta < 1/ 2$ is arbitrary. Then, the Proposition follows
from the definition of $E^*(\ep)$ and (\ref{e:theta_n_Gaussian}) as in the proof of Theorem 
\ref{t:converse}. The Corollary results by a straightforward calculation. It remains to verify that
(\ref{e:Entropy_boundsOnP3}) holds  for
$\theta$ in (\ref{e:theta_n_Gaussian}). 
For $B \subsetneq \cM, B \neq \emptyset$, let $g_B$ denote the density 
of the Gaussian rv $X_B^{(n)}$.
From the AEP for Gaussian rvs \cite[equation (47)]{CovPom89} (see also \cite{BobMad11}),
\begin{align}
&\mathrm{P}\bigg(\left|-\frac{1}{n}\log g_B\left(X_B^{(n)}\right) - \frac{1}{n}h\left(X_B^{(n)}\right)\right| > \tau,
\nonumber\\
&\hspace*{3cm} \text{ for some } \emptyset\neq B\subseteq \cM\bigg) 
\nonumber\\
&< 2^m \exp (-c(\tau)n), \quad \tau >0,
\label{e:AEP_Gaussian}
\end{align}
where $h$ denotes differential entropy and $c(\tau) >0$ is a positive constant that does not
depend on $n$. Since 
\begin{align}
\frac{d\,\mathrm{P}}{d\,\tilde{\mathtt{P}}} = \frac{g_\cM}{\prod_{i=1}^{|\pi|}g_{\pi_i}}, \qquad \mathrm{P}\text{ a.s.}
\nonumber
\end{align}
and 
\begin{align}
&h\left(X_\cM^{(n)}\right) = \frac{1}{2}\log (2\pi e)^{mn}|\Sigma^{(n)}|,
\nonumber\\
& h\left(X_{\pi_i}^{(n)}\right)= \frac{1}{2}\log (2\pi e)^{|\pi_i|n}\big|\Sigma^{(n)}_{\pi_i}\big|, \qquad 1 \leq i \leq |\pi|,
 \nonumber
\end{align}
using the upper and lower bounds from (\ref{e:AEP_Gaussian}) that hold with 
significant probability for all $n$ sufficiently large, 
we get that (\ref{e:Entropy_boundsOnP3}) holds with
$\theta$ as in (\ref{e:theta_n_Gaussian}), for $0<\delta < 1/ 2$.\qed

As an application of the Corollary above, we establish a new strong converse for SK
capacity when the underlying rvs $X_\cM^{(n)}$ are i.i.d. Gaussian in $n$; for this model, the 
SK capacity was established in \cite{NitNar12}. The notions of $\ep$-achievable SK rate, $\ep$-SK
capacity $C(\ep)$ and SK capacity $C$ are as in Definition \ref{d:SK_cap}, with condition 
(\ref{e:SK_rate}) replaced by
\begin{align}
\mathtt{range}(K) = \{1 , ..., \lfloor\exp (nR)\rfloor\},
\label{e:range_SK}
\end{align}
which rules out such rvs $K$ as take
infinitely many values.
\begin{proposition}
When $X_\cM^{(n)}$ is i.i.d. in $n$ with $X_\cM \sim \cN(\mathbf{0}, \Sigma)$,
\begin{align}
C(\ep) = \min_\pi \frac{1}{2(|\pi| -1)} \log\frac{\prod_{i=1}^{|\pi|}
\big|\Sigma_{\pi_i}\big|}{|\Sigma|}, \quad 0 < \ep <1.
\label{e:Gaussian_capacity}
\end{align}
\end{proposition}
{\it Proof.} That $C(\ep)$ is no smaller than the right-side of (\ref{e:Gaussian_capacity}) follows from the achievability proof in \cite{NitNar12}.

The proof of the reverse inequality is along the lines of the proof of Theorem \ref{t:strong_converse_SK_cap}
and is obtained upon replacing the upper bound (\ref{e:ub_card_K}) by 
\begin{align}
\frac{\rho}{n}\log\frac{n}{\rho} + 2\rho R,
\nonumber
\end{align}
and noting that Lemma \ref{l:equivalence} can be extended straightforwardly to an
arbitrary rv $V$ (with the explicit summations in the proof of that Lemma written 
as expectations), provided that the rv $U$ is finite-valued.\qed

{\it Proof of Theorem \ref{t:very_general_converse}.} In the manner of the proof of Theorem \ref{t:general_converse}, it suffices to identify measurable sets $\cI_0$ and 
$\cL(\bi) \subseteq \cL$, $\bi \in \cI_0$, such that (\ref{e:L(i)_lb_P})-(\ref{e:I0_lb_P})
are satisfied. Below we generalize appropriately the steps 1-3 in the proof of Theorem \ref{t:general_converse}. 

\noindent {\it Step 1.} The following claim is an extension of Lemma \ref{l:indep_condF}.

\begin{lemma}\label{l:indep_condF3}
Given measurable sets $A_i \in \sigma_i$, $1\leq i \leq k$, for  $\tilde{\mathrm{P}}$ in 
(\ref{e:tP_def3}), 
\begin{align}
\tP{Y_1, ..., Y_k\mid \bF}{A_1\times...\times A_k\mid \bi} &= \prod_{j=1}^k \tP{Y_j \mid \bF}{A_j\mid \bi}, 
\nonumber\\
&\hspace*{1.7cm} \bPP{\bF} \text{ a.s. in }\bi,
\label{e:indep_tp}
\end{align}
where $\tPP{Y_1,..., Y_k\mid \bF}$ is the regular conditional probability on $(\cY^k, \sigma^k)$
conditioned on $\sigma(\bF)$.
\end{lemma}
\noindent The proof uses the interactive property of the communication and is relegated to the
Appendix.

\noindent {\it Step 2.} Next, we identify the set $\cI_0$. The following technical observation will be used.

\begin{lemma}\label{l:tP_absolute_cont}
For every $A_0 \in \sigma^k$ such that 
\begin{align}
\frac{d\, \mathrm{P}}{d\, \tilde{\mathrm{P}}}(y^k) > 0,\qquad y^k \in A_0,
\label{e:positive_RN}
\end{align}
it holds that
\begin{align}
\tP{Y_1,..., Y_k\mid \bF}{A_0\mid \bi} = \frac{d\, \bPP{\bF}}{d\, \tPP{\bF}}(\bi) \int_{A_0} 
\frac{d\,Q_\bi}{d\, \mathrm{P}}\,  d\,\tilde{\mathrm{P}},\quad \tPP{\bF} \text{ a.s. in }\bi
\label{e:tP_absolute_continuity}
\end{align}
\end{lemma} 
\noindent The proof is given in the Appendix. Denoting by $\cT_0$ the set 
$\left\{y^k \in \cY^k: 0 < \frac{d\, \mathrm{P}}{d\, \tilde{\mathrm{P}}}(y^k) \leq \theta \right\}$,
let 
\begin{align}
A_l = \cT_0 \cap \left\{ y^k: L_j\big(y_j, \bF(y^k)\big) = 
L(y^k) = l, 1\leq j \leq k\right\}, \qquad l \in \cL.
\nonumber
\end{align}
Then, for $A_{l, \bi} \triangleq A_l \cap \bF^{-1}(\bi)$, 
(\ref{e:assumption_general2i}), (\ref{e:assumption_general2ii}) and Lemma \ref{l:tP_absolute_cont} imply
\begin{align}
\tP{Y_1, ..., Y_k\mid \bF}{A_{l, \bi}\mid \bi} = 
\frac{d\, \bPP{\bF}}{d\, \tPP{\bF}}(\bi) 
\int_{A_{l,\bi}} \frac{d\, Q_\bi}{d\, \mathrm{P}} \, d\,\tilde{\mathrm{P}},\quad \tPP{\bF}\text{ a.s. in }\bi. 
\label{e:tP_absolute_continuity2}
\end{align}
Below we restrict attention to the set of values of $\bF$ for which (\ref{e:tP_absolute_continuity2}) holds
for every $l \in \cL$; this set has $\tPP{\bF}$ measure $1$ by (\ref{e:tP_absolute_continuity})
since the set $\cL$ is countable. 
Proceeding along the lines of the proof of Theorem \ref{t:general_converse}, we define $\cI_1$  
as the set of those $\bi$ for which
\begin{align}
\bP{Y_1, ..., Y_k\mid \bF}{A_{l, \bi}\mid \bi} \geq 1 - \sqrt{\ep + \delta}.
\label{e:lb_P_Ali3}
\end{align}
Since $L$ is an $\ep$-CR from $\bF$, it follows from (\ref{e:Entropy_boundsOnP3}), the fact that $$\bPr{\left\{ y^k: \frac{d\, \mathrm{P}}{d\, \tilde{\mathrm{P}}}(y^k) = 0\right\}} = 0,$$ and
by a reverse Markov inequality, that
\begin{align}
\bP{\bF}{\cI_1} \geq 1 - \sqrt{\ep+\delta}.
\label{e:lb_I1_P3}
\end{align}
Furthermore, for the set $\cI_2$ of values $\bi$ of $\bF$ satisfying 
\begin{align}
\frac{d\, \bPP{\bF}}{d\, \tPP{\bF}}(\bi) \geq \delta,
\label{e:I2_def3}
\end{align}
it holds that 
\begin{align}
\bP{\bF}{\cI_2}\geq 1 -\delta,
\label{e:lb_I2_P3}
\end{align}
since
\begin{align}
\int_{\cI_2^c} d\, \bPP{\bF} &= \int_{\cI_2^c} \frac{d\, \bPP{\bF}}{d\, \tPP{\bF}} d\, \tPP{\bF}
\nonumber \\
&< \delta.
\nonumber
\end{align}
Define $\cI_0 = \cI_1 \cap \cI_2$; (\ref{e:I0_lb_P}) follows from (\ref{e:lb_I1_P3})
and (\ref{e:lb_I2_P3}).

\noindent {\it Step 3.} Since Lemma \ref{l:card_largePset} (i) applies to a countable set $\cU = \cL$, defining the nonnegative measure $\mu$ on $\cL$ as in (\ref{e:mu_def1}) for each $\bi \in \cI_0$ and using (\ref{e:lb_P_Ali3}), the sets $\cL(\bi)$ obtained in (\ref{e:L(i)_lb_P_proof})-(\ref{e:lb_P_LgivenF}) satisfy (\ref{e:L(i)_lb_P}). Also, condition (\ref{e:L(i)_ub_Card}) will follow from (\ref{e:L(i)_ub_Renyi}) upon showing 
that
\begin{align}
\exp\left(H_\alpha(\mu)\right) \leq \left(\frac{\theta}{\delta}\right)^{\frac{1}{k-1}}.
\label{e:ub_RenyiEntropy3}
\end{align}

To do so, denote by  $A_{l,\bi}^j$ the projection of $A_{l,\bi}$ along the $j$th coordinate, $1\leq j \leq k$. As before, the sets $A_{l,\bi}^j$ are disjoint across $l \in \cL$. Then, H\"older's inequality \cite{HarLitPol52} implies that
\begin{align}
1 &\geq \prod_{j=1}^k \left[\sum_{l \in \cL} \tP{Y_j \mid \bF}{A_{l,\bi}^j \mid \bi}\right]
\nonumber\\
&\geq \left[\sum_{l \in \cL} \left(\prod_{j=1}^k \tP{Y_j \mid \bF}{A_{l,\bi}^j \mid \bi}^{\frac{1}{k}}\right)\right]^k
\nonumber\\
&= \left[\sum_{l \in \cL}  \tP{Y_1, ..., Y_k \mid \bF}{A_{l,\bi}^1\times...\times A_{l,\bi}^k \mid \bi}^{\frac{1}{k}}\right]^k,
\label{e:lb3i}
\end{align}
where the previous step uses Lemma \ref{l:indep_condF3}. The right-side of (\ref{e:lb3i}) is bounded below by 
\begin{align}
\left[\sum_{l \in \cL}  \tP{Y_1, ..., Y_k \mid \bF}{A_{l,\bi} \mid \bi}^{\frac{1}{k}}\right]^k,
\nonumber
\end{align}
since $A_{l,\bi} \subseteq A_{l,\bi}^1\times...\times A_{l,\bi}^k $, which by (\ref{e:tP_absolute_continuity2}) equals
\begin{align}
\left[\sum_{l \in \cL}  \left(\frac{d\, \bPP{\bF}}{d\, \tPP{\bF}}(\bi) \int_{A_{l,\bi}} 
\frac{d\,Q_\bi}{d\, \mathrm{P}}  d\,\tilde{\mathrm{P}}\right)^{\frac{1}{k}}\right]^k.
\nonumber
\end{align}
From the definition of the set $\cI_2$ in (\ref{e:I2_def3}), the expression above exceeds
\begin{align}
\left[\sum_{l \in \cL}  \left(\delta\int_{A_{l,\bi}} 
\frac{d\,Q_\bi}{d\, \mathrm{P}}  d\,\tilde{\mathrm{P}}\right)^{\frac{1}{k}}\right]^k,
\nonumber
\end{align}
which is the same as 
\begin{align}
\left[\sum_{l \in \cL}  \left(\delta\int_{A_{l,\bi}} 
\frac{d\, Q_\bi}{d\, \mathrm{P}} \frac{d\, \mathrm{P}/  d\,\tilde{\mathrm{P}}}{d\, \mathrm{P}/  d\,\tilde{\mathrm{P}}} d\,\tilde{\mathrm{P}}\right)^{\frac{1}{k}}\right]^k.
\label{e:lb3ii}
\end{align}
Since $A_{l, \bi} \subseteq \cT_0$, the sum in (\ref{e:lb3ii}) is bounded below further by
\begin{align}
&\left[\sum_{l \in \cL}  \left(\frac{\delta}{\theta}\int_{A_{l,\bi}} 
\frac{d\, Q_\bi}{d\, \mathrm{P}} \frac{d\, \mathrm{P}}{d\, \tilde{\mathrm{P}}}   d\,\tilde{\mathrm{P}}\right)^{\frac{1}{k}}\right]^k
\nonumber \\
&= \frac{\delta}{\theta}\left[\sum_{l \in \cL}  \left(\int_{A_{l,\bi}} 
d\, Q_\bi\right)^{\frac{1}{k}}\right]^k
\nonumber \\
&= \frac{\delta}{\theta}\left[\sum_{l \in \cL}  
\bP{Y_1, ..., Y_k\mid \bF}{A_{l,\bi}\mid \bi}^{\frac{1}{k}}\right]^k.
\nonumber
\end{align}
Combining the observations above from (\ref{e:lb3i}) onward, we have 
\begin{align}
\frac{\theta}{\delta} \geq \left[\sum_{l \in \cL}  
\bP{Y_1, ..., Y_k\mid \bF}{A_{l,\bi}\mid \bi}^{\frac{1}{k}}\right]^k,
\nonumber
\end{align}
which is the same as (\ref{e:ub_RenyiEntropy3}) with $\alpha= 1/k$.\qed


\section{Discussion}\label{s:Discussion}

\subsection{General lossless source coding theorem} Our Lemma \ref{l:card_largePset}
relating the cardinalities of large probability sets to R{\'e}nyi entropy played a material role
in the converse proofs. It is also of independent interest, and can be interpreted as a source 
coding result for a general source with finite alphabet $\cU$. Furthermore, it leads to the 
following asymptotic result. 

Consider a sequence of probability measures $\mu_n$ on finite sets $\cU_n$, $n \geq 1$.
For $0< \delta <1$, $R$ is a $\delta$-achievable (block) source coding rate if there exists sets 
$\cV_n \subseteq \cU_n$ satisfying
\begin{align}
\mu_n(\cV_n) \geq 1-\delta, 
\nonumber
\end{align}
for all $n$ sufficiently large, and
\begin{align}
\limsup_n \frac{1}{n}\log |\cV_n| \leq R.
\nonumber
\end{align}
The optimum source coding rate $R^*(\delta)$ is the infimum of all such 
$\delta$-achievable rates. 
\begin{proposition}\label{p:general_source_coding}
For each $0< \delta < 1$, 
\begin{align}
\lim_{\alpha \downarrow 1} \,\,\limsup_n \frac{1}{n} H_\alpha(\mu_n) \leq R^*(\delta) \leq \lim_{\alpha\uparrow 1} \,\,\limsup_n \frac{1}{n}H_\alpha(\mu_n).
\label{e:lb_ub_R*(delta)}
\end{align}
\end{proposition}
\begin{corollary}
If $\mu_n$ is an i.i.d. probability measure on $\cU_n = \cU\times ... \times \cU$, then 
$$R^*(\delta) = H(\mu_1), \qquad 0 < \delta < 1.$$
\end{corollary}
\noindent {\it Proof.} The Proposition is a direct consequence of Lemma \ref{l:card_largePset} upon
taking appropriate limits in (\ref{e:ub_card_largePset}) and (\ref{e:lb_card_largePset}) with $\cU_n$
in the role of $\cU$. The Corollary follows since for i.i.d. $\mu_n$, 
$$H_\alpha(\mu_n) = nH_\alpha(\mu_1) \text{ and } \lim_{\alpha\rightarrow 1} H_\alpha(\mu_1) = H(\mu_1).$$
\qed

Note that the Corollary above is proved without recourse to
the AEP. Moreover, it contains 
a strong converse for the lossless coding theorem for an 
i.i.d. source. In general, Proposition \ref{p:general_source_coding} implies
a strong converse whenever the lower and upper bounds for $R^*(\delta)$
in (\ref{e:lb_ub_R*(delta)}) coincide. This implication is a special case of 
a general source coding result in \cite[Theorem 1.5.1]{Han03}, \cite{HanVer93}, where it was shown
that a strong converse holds iff for rvs $U_n$ with pmfs $\mu_n$, the
``lim-inf" and ``lim-sup" of $Z_n = \frac{1}{n}\log\frac{1}{\mu_n(U_n)}$
in $\mu_n$-probability coincide, i.e.,
\begin{align}
&\sup\left\{ \beta : \lim_n \mu_n(Z_n < \beta) =0\right\} 
\nonumber\\
&= \inf\left\{ \beta : \lim_n \mu_n(Z_n >\beta) =0\right\}.
\label{e:liminf=limsup}
\end{align}
In fact, a straightforward calculation shows that the lower and upper bounds 
for $R^*(\delta)$ in (\ref{e:lb_ub_R*(delta)}) are admissible choices of
$\beta$ on the left- and right-sides of (\ref{e:liminf=limsup}), respectively.

\subsection{General Models}
The description of the optimum query exponent in Definition \ref{d:query_exponent} can be refined 
to display an explicit dependence on $\ep^\prime$. Let $E^*(\ep, \ep^\prime)$
denote the optimum query exponent for fixed $0 < \ep, \ep^\prime < 1$.
Our proofs establish $E^*(\ep, \ep^\prime)$ equals the right side of (\ref{e:E*_char})
for $\ep^\prime < (1- \sqrt{\ep})^2$ (see (\ref{e:32a})). For $\ep^\prime> 1- \ep$,
as suggested by a reviewer, the following construction of $L$ renders $E^*(\ep, \ep^\prime)$
unbounded: Choose $L = 0$ with probability $(1-\ep)$ and uniformly distributed 
on a sufficiently large set with probability $\epsilon$. For the remaining values of $\ep, \ep^\prime$,
$E^*(\ep, \ep^\prime)$ is not known. 

A less restrictive model for querying than that in Section \ref{s:Main Results} can be considered, 
allowing general 
queries with binary answers. Such a query strategy can be represented 
as a search on a binary tree whose leaves correspond to the 
values of the CR $L$. The query strategies considered in this paper 
correspond to the case where the search tree is a path with leaves
attached to each node. For a general tree model, our results can 
be adapted to show that the maximum number of queries that can 
be inflicted on a querier grows only linearly in $n$ at a rate that is equal to 
the expression for $E^*$ in (\ref{e:E*_char}).

We remark also that allowing randomness at the terminals
in $\cM$ for interactive communication and CR recovery,
does not improve the optimum query exponent. Such 
randomization is described by mutually independent rvs 
$W_1, ..., W_m$, where each $W_i$ is distributed uniformly
on the (finite) set $\left\{1, ..., w_i\right\}$, and the rvs 
$W_1, ..., W_m$ are independent of $X_\cM^n$. The claim
of the remark is seen from the converse result in 
Theorem \ref{t:general_converse2}. Indeed, the assertion (\ref{e:q_adversary2})
of Theorem \ref{t:general_converse2} remains unchanged upon 
replacing $Y_i$ by $\left(Y_i, W_i\right)$, $i \in \cM$, 
$\theta_0$ by $\theta_0\left(\prod_{i \in \cM}w_i\right)$,
and $\theta_{B^c}$ by $\theta_{B^c}\left(\prod_{i \in B^c}w_i\right)$, $B \in \cB$;
and observing that in (\ref{e:theta_def}), the $w_i$- terms cancel
in the numerator and the denominator.

Finally, Lemma \ref{l:equivalence}, which considered rvs $U, V$,
can be used to characterize the optimum query exponent $\Gamma^*$
for a family of finite-valued rvs $\left\{U_n, V_n\right\}_{n=1}^\infty$
with associated probability measures $\left\{\bPP{n}\right\}_{n=1}^\infty$
(which are not necessarily consistent). Here, $\Gamma^*$ is described 
analogously as $E^*$ in Definition \ref{d:query_exponent}. An application
of Lemma \ref{l:equivalence} yields that
\begin{align}
 \Gamma^* &\geq \bPP{n}\text{-}\liminf_n \frac{-\log \bP{U_n\mid V_n}{U_n\mid V_n}}{n} 
  \nonumber\\
\Gamma^* &\leq\, \bPP{n}\text{-}\limsup_n \frac{-\log \bP{U_n\mid V_n}{U_n\mid V_n}}{n}
\nonumber
\end{align}
where the first and second limits above equal, respectively, the left- and right-sides 
of (\ref{e:liminf=limsup}) with $\mu_n = \mathrm{P}_n$ and
$$Z_n = \frac{-\log \bP{U_n\mid V_n}{U_n\mid V_n}}{n}.$$


\section*{Appendix}
\setcounter{equation}{0}
\renewcommand{\theequation}{A\arabic{equation}}

\begin{center}
\noindent{\it Proof of Lemma \ref{l:indep_condF3}.} 
\end{center}

For $1\leq l \leq r$, $1\leq j \leq k$, denote by $\Phi_{lj}$ the interactive communication preceding $F_{lj}$, by $\bF_{lj}$ the rv $\left(F_{lj}, \Phi_{lj}\right)$, and by $\bi_{lj}$ a realization of $\bF_{lj}$. Without loss of generality,
we choose a version of $\tPP{Y^k\mid \bF}$ that satisfies
\begin{align}
\tP{Y^k\mid \bF_{lj}}{\bF_{lj}^{-1}(\bi_{lj})^c \mid \bi_{lj}} = 0, \quad  \tPP{\bF_{lj}}\text{ a.s.,}
\label{e:A1}
\end{align}
for all $1\leq l \leq r, 1\leq j \leq k$. The following property of interactive communication
is pivotal to our proof:  For each 
$i_{lj}^-$, $\Phi_{lj}^{-1}(i_{lj}^-)$ is a product set, i.e., 
\begin{align}\nonumber
\Phi_{lj}^{-1}(i_{lj}^-) = A_1^\prime\times ... \times A_k^\prime, \quad A_j^\prime \in \sigma_j,\,\, 1\leq j \leq k. 
\end{align}
We prove the claim by induction upon observing that $\tPP{Y^k\mid \bF_{lj}}$ can be obtained by conditioning
$\tPP{Y^k \mid \Phi_{lj}}$ on the rv $F_{lj}$.

Formally,
denote by $\sigma^k(i_{lj}^-) = \sigma_1(i_{lj}^-)\times ... \times \sigma_k(i_{lj}^-)$ the $\sigma$-field induced by $\sigma^k$ on $A_1^\prime \times... \times A_k^\prime$, and by $\sigma\left(F_{lj}(\cdot, i^-_{lj})\right)$ the smallest sub-$\sigma$-field of $\sigma^k(i_{lj}^-)$ with respect to which $F_{lj}$ is measurable (for $i^-_{lj}$ fixed).
Using (\ref{e:A1}), we choose a version of $\tPP{Y^k\mid \bF}$ such that for each $1\leq l \leq r$ and $1\leq j \leq k$, $\tP{Y^k\mid \bF_{lj}}{\cdot\mid \bi_{lj}}$ is the regular conditional probability on the probability space
$$\left(A_1^\prime \times ... \times A_k^\prime,\,\, \sigma^k(i_{lj}^-),\,\, \tP{Y^k \mid \Phi_{lj}}{\cdot \mid i_{lj}^-}\right)$$
conditioned on
$\sigma\left(F_{lj}(\cdot, i^-_{lj})\right)$. Specifically,
\begin{align}
&\tP{Y^k\mid \bF_{lj}}{A\mid \bi_{lj}} 
\nonumber\\
&= \bE{\tP{Y^k \mid \Phi_{lj}}{\cdot \mid i_{lj}^-}}{\mathbf{1}_A\mid \sigma\left(F_{lj}(\cdot, i^-_{lj})\right)}(i_{lj}), 
\quad A\in \sigma^k,
\label{e:A2}
\end{align}
where the underlying $\sigma$-field for the conditional expectation is $\sigma^k(i_{lj}^-)$. 
For this version of $\tPP{Y^k\mid \bF}$, we show below that if (\ref{e:indep_tp}) 
holds with $\Phi_{lj}$ in the role of $\bF$, then it holds with $\bF_{lj}$ in the role of $\bF$. Lemma \ref{l:indep_condF3} then follows by induction since (\ref{e:indep_tp}) 
holds with $\bF = \emptyset$. 

It remains to prove the assertion above. To that end, 
for $B \in \cF_{lj}$, denote by 
$F_{lj}^{-1}\left(B, i_{lj}^-\right)$  the set 
$$\left\{y_j \in \cY_j : F_{lj}\left(y_j, i_{lj}^-\right)\in B\right\}.$$
With an abuse of notation, we do not distinguish between the sets $F_{lj}^{-1}\left(B, i_{lj}^-\right)$ and its cylindrical extension
$$\cY_1\times ... \times F_{lj}^{-1}\left(B, i_{lj}^-\right) \times... \times\cY_k.$$
Then, using the notation $\tilde{Q}_{i_{lj}^-}$
and $\tilde{Q}^t_{i_{lj}^-}, 1 \leq t \leq k$,
for the probability
measures $\tP{Y^k\mid \Phi_{lj}}{\cdot\mid i_{lj}^-} $
and $\tP{Y_t\mid \Phi_{lj}}{\cdot\mid i_{lj}^-} $, $1 \leq t \leq k$, respectively, 
our induction hypothesis states
\begin{align}\label{e:A_ih}
\tilde{Q}{i_{lj}^-}(A_1 \times .... \times A_k) = \prod_{t=1}^k \tilde{Q}{i_{lj}^-}(A_t), \quad A_t\in \sigma_t, \,\,1\leq t \leq k.
\end{align}
It follows that
\begin{align}
&\int_{F_{lj}^{-1}(B, i_{lj}^-)} \mathbf{1}_{A_1\times ... \times A_k} \, d\, \tilde{Q}_{i_{lj}^-}
\nonumber\\
&= \int_{F_{lj}^{-1}(B, i_{lj}^-)} \mathbf{1}_{A_1\cap A_1^\prime\times ... \times A_k\cap A_k^\prime} \, d\, \tilde{Q}_{i_{lj}^-}
\nonumber\\
&= \left[\prod_{t\neq j}\int \mathbf{1}_{A_t\cap A_t^\prime}\,d\, \tilde{Q}^t_{i_{lj}^-}\right]
 \int_{F_{lj}^{-1}(B, i_{lj}^-)}\mathbf{1}_{A_j\cap A_j^\prime}\,d\, \tilde{Q}^j_{i_{lj}^-},
\label{e:A3}
\end{align}
where the first equality uses (\ref{e:A1}) and the second uses (\ref{e:A_ih}). Defining
\begin{align}
P_{lj}^t(A) \triangleq &
\bE{\tilde{Q}{i_{lj}^-}^t}{\mathbf{1}_A\mid \sigma\left(F_{lj}(\cdot, i^-_{lj})\right)},
\nonumber\\
& \hspace*{3cm} A\in \sigma_t(i_{lj}^-), \,\,1\leq t\leq k,
\nonumber
\end{align}
we have from (\ref{e:A3}) that
\begin{align}
&\int_{F_{lj}^{-1}(B, i_{lj}^-)} \mathbf{1}_{A_1\times ... \times A_k} \, d\, \tilde{Q}_{i_{lj}^-}
\nonumber\\
&= \left[\prod_{t\neq j}\int P_{lj}^t(A_t \cap A_t^\prime)\,d\, \tilde{Q}^t_{i_{lj}^-}\right]\times
\nonumber\\
&\hspace*{2cm} \int_{F_{lj}^{-1}(B, i_{lj}^-)} P_{lj}^j(A_j \cap A_j^\prime)\,d\, \tilde{Q}^j_{i_{lj}^-}
\nonumber\\
&=\int_{F_{lj}^{-1}(B, i_{lj}^-)} \prod_{t=1}^k P^t_{lj}(A_t \cap A_t^\prime) \, d\, 
\tilde{Q}_{i_{lj}^-},
\nonumber
\end{align}
where the second equality uses (\ref{e:A_ih}).
Thus, by (\ref{e:A2}),
\begin{align}
&\tP{Y^k \mid \bF_{lj}}{A_1\times ... \times A_k\mid \bi_{lj}}
\nonumber\\
 &= \prod_{t=1}^k P^t_{lj}(A_t \cap A_t^\prime), \quad \tPP{\bF_{lj}}\text{ a.s. in } \bi_{lj}.
\label{e:A4}
\end{align}
Since by (\ref{e:A1}) $P^t_{lj}(A_t^\prime) = 1$, $1\leq t \leq k$, it follows from (\ref{e:A4}) that
\begin{align}
&P_{lj}^t(A_t) 
\nonumber\\
&=P_{lj}^t(A_t\cap A_t^\prime) 
\nonumber\\
&=\tP{Y^k \mid \bF_{lj}}{A_1^\prime\times ...\times A_{t-1}^\prime \times A_t \times A_{t+1}^\prime\times ...\times A_k^\prime\mid \bi_{lj}}
\nonumber\\
&= \tP{Y_t\mid \bF_{lj}}{A_t\mid \bi_{lj}}.
\nonumber
\end{align}
The previous observation, along with (\ref{e:A4}), implies that (\ref{e:indep_tp}) holds with $\bF_{lj}$ in the role $\bF$.\qed

\begin{center}
{\it Proof of Lemma \ref{l:tP_absolute_cont}.}
\end{center}

It suffices to show that the right-side of (\ref{e:tP_absolute_continuity}) 
constitutes a version of $\bE{\tilde{\mathrm{P}}}{\mathbf{1}_{A_0} \mid \sigma(\bF)}$, i.e.,
\begin{align}
\int_{\bF^{-1}(B)}  \mathbf{1}_{A_0} d\,\tilde{\mathrm{P}} =
\int_{B} \left( \int_{A_0} \frac{d\, \bPP{\bF}}{d\, \tPP{\bF}}(z) 
\frac{d\,Q_z}{d\, \mathrm{P}}  d\,\tilde{\mathrm{P}}\right)\tP{\bF}{d\,z},
\label{e:tP_abs_cont4}
\end{align}
for every set $B$ in the range $\sigma$-field of $\bF$. To show that, we note for
every $A\in \sigma^k$ that
\begin{align}
\int_{\bF^{-1}(B)}  \mathbf{1}_{A} d\,\mathrm{P} &= \int_B \bP{Y\mid \bF}{A\mid z} \bP{\bF}{d\,z}
\nonumber\\
&=\int_B \left(\int_A\frac{d\, Q_z}{d\, \mathrm{P}}\, d\,\mathrm{P}\right) \bP{\bF}{dz},
\label{e:a1}
\end{align}
where the previous step uses the assumption (\ref{e:assumption_general}). Using Fubini's and Tonelli's theorems 
to interchange the order of integrals in (\ref{e:a1}), we get
\begin{align}
\int_{\bF^{-1}(B)}  \mathbf{1}_{A}\, d\,\mathrm{P}
&= \int_A\left(\int_B\frac{d\,Q_z}{d\, \mathrm{P}}\, \bP{\bF}{dz}\right) d\, \mathrm{P},
\nonumber\\
&= \int_A\mathbf{1}_{\bF^{-1}(B)}\, d\,\mathrm{P},
\nonumber
\end{align}
which further implies
\begin{align}
\mathbf{1}_{\bF^{-1}(B)} = \int_B\frac{d\, Q_z}{d\, \mathrm{P}}  \bP{\bF}{dz}, \qquad \mathrm{P}\text{ a.s.},
\label{e:a2}
\end{align}
since the set $A \in \sigma^k$ was arbitrary. Next, for every $B$ in the range $\sigma$-field of $\bF$, it follows from (\ref{e:a2}) and (\ref{e:positive_RN}) that
\begin{align}
 \int_{\bF^{-1}(B)}\mathbf{1}_{A_0}d\,\tilde{\mathrm{P}} &= \int_{A_0}\mathbf{1}_{\bF^{-1}(B)} d\,\tilde{\mathrm{P}}
\nonumber\\
&= \int_{A_0}\frac{1}{d\, \mathrm{P}/  d\, \tilde{\mathrm{P}}}\mathbf{1}_{\bF^{-1}(B)} d\,\mathrm{P}
\nonumber\\
&= \int_{A_0}\frac{1}{d\, \mathrm{P}/  d\, \tilde{\mathrm{P}}}\int_B\frac{d\, Q_z}{d\, \mathrm{P}}  \bP{\bF}{dz} d\,\mathrm{P}
\nonumber\\
&= \int_{A_0}\int_B\frac{d\, Q_z}{d\, \mathrm{P}}  \bP{\bF}{dz} d\,\tilde{\mathrm{P}}.
\label{e:a3}
\end{align}
The claim (\ref{e:tP_abs_cont4}) follows upon interchanging the order of integrals in (\ref{e:a3}).
\qed

\section*{Acknowledgement}
The authors thank Mokshay Madiman for useful pointers
to the work on the AEP for log-concave distributions and related
literature \cite{BobMad11}, \cite{CovPom89}, which were instrumental
in our
results for Gaussian rvs in Section \ref{s:general_converse}. Thanks are also 
due to an anonymous referee for suggesting the refined definition $E^*(\ep, \ep^\prime)$
discussed in Section \ref{s:Discussion}-B.



\end{document}